\documentclass[12pt]{article}
\usepackage{epsfig,amsfonts,amsmath,array,mathrsfs}
\newcommand{\pbs}[1]{\let\temp=\\#1\let\\=\temp}
\renewcommand{\theequation}{\thesection.\arabic{equation}}
\def\be{\begin{equation}}\def\ee{\end{equation}}
\def\cvp{\raise 2pt\hbox{,}}

 \def\Tr{\mathop{\rm
Tr}\nolimits}

\def\re{\mathop{\rm Re}\nolimits} 
\def\diag{\mathop{\rm diag}\nolimits}

\def\d{{\rm d}}
\def\nn{{\cal N}}

 \def\uN{{\rm U}(N)} 
  
\def\La{\Lambda}
\def\wmic{W_{\text{mic}}}

\def\a{\boldsymbol{a}}
\def\g{\boldsymbol{g}}
 \def\m{\boldsymbol{m}}

\def\uR{\text{U}(1)_{\text R}} \def\u{\text{U}(1)}

\def\eps{\epsilon}
\def\vevas#1{\langle\a|#1|\a\rangle}
\def\vevab#1{\bigl\langle\a\big|#1\big|\a\bigr\rangle}
\def\cpart{\vec{\mathsf k}}
\def\cl{\text{cl}} 
\def\plb#1#2#3{{\it Phys.\ Lett.\ }{\bf B #1} (#2) #3}
\def\npb#1#2#3{{\it Nucl.\ Phys.\ }{\bf B #1} (#2) #3}
\def\npps#1#2#3{{\it Nucl.\ Phys.\ Proc.\ Suppl.\ }{\bf #1} (#2) #3}

\def\jhep#1#2#3{{\it JHEP\ }{\bf #1} (#2) #3}
\def\prd#1#2#3{{\it Phys.\ Rev.\ }{\bf D #1} (#2) #3}

\def\atmp#1#2#3{{\it Adv.\ Theor.\ Math.\ Phys.\ }{\bf #1} (#2) #3}
\def\cmp#1#2#3{{\it Comm.\ Math.\ Phys.\ }{\bf #1} (#2) #3}
\def\pr#1#2#3{{\it Phys.\ Rep.\ }{\bf #1} (#2) #3}

\def\rmp#1#2#3{{\it Rev.\ Mod. Phys. }{\bf #1} (#2) #3}

\begin{document}
%
%
\pagestyle{empty}
{\parskip 0in

\hfill LPTENS-07/34

\hfill arXiv:0708.1410 [hep-th]}

\vfill
\begin{center}
{\LARGE Glueball operators and}

\medskip

{\LARGE the microscopic approach to $\nn=1$ gauge theories}

\vspace{0.4in}

Frank \textsc{Ferrari}, 
Stanislav \textsc{Kuperstein} and Vincent \textsc{Wens}
\\
\medskip
{\it Service de Physique Th\'eorique et Math\'ematique\\
Universit\'e Libre de Bruxelles and International Solvay Institutes\\
Campus de la Plaine, CP 231, B-1050 Bruxelles, Belgique
}\\
\smallskip
{\tt frank.ferrari@ulb.ac.be, skuperst@ulb.ac.be, vwens@ulb.ac.be}
\end{center}
\vfill\noindent

We explain how to generalize Nekrasov's microscopic approach to
$\nn=2$ gauge theories to the $\nn=1$ case, focusing on the typical
example of the $\uN$ theory with one adjoint chiral multiplet $X$ and
an arbitrary polynomial tree-level superpotential $\Tr W(X)$. We
provide a detailed analysis of the generalized glueball operators and
a non-perturbative discussion of the Dijkgraaf-Vafa matrix model and
of the generalized Konishi anomaly equations. We compute in particular
the non-trivial quantum corrections to the Virasoro operators and
algebra that generate these equations. We have performed explicit
calculations up to two instantons, that involve the next-to-leading
order corrections in Nekrasov's $\Omega$-background.

\vfill

\medskip
%
\begin{flushleft}
\today
\end{flushleft}
\newpage\pagestyle{plain}
\baselineskip 16pt
\setcounter{footnote}{0}

\section{Introduction}
\setcounter{equation}{0}

Recently, a very general strategy to derive non-perturbative exact
results in $\nn=1$ gauge theories from a microscopic point of view was
explained \cite{mic1}. The starting point is to consider the gauge
theory path integral with arbitrary boundary conditions at infinity. A
microscopic quantum effective superpotential $\wmic$ can be derived as
a function of the boundary conditions. This effective superpotential
has two fundamental properties. First, it can always be computed
exactly in a semi-classical instanton framework by choosing the
boundary conditions appropriately and then performing suitable
analytic continuations. Second, the stationary points of $\wmic$
describe \emph{all} the quantum vacua of the theory, including the
strongly coupled confining vacua. A direct procedure for solving the
theory in the chiral sector from microscopic instanton calculations
then follows. In particular, the full power of Nekrasov's technology
\cite{nekrasova}, which itself was the crowning achievement of many
years of developments in instanton calculus
\cite{insta,instb,instc,instrev} and which was successfull in solving
$\nn=2$ gauge theories \cite{nekrasovb}, can be applied to the realm
of $\nn=1$ gauge theories, generalizing useful early work
\cite{fucito}.

The basic example on which to apply these ideas is the $\nn=1$ theory
with gauge group $\uN$, one adjoint chiral superfield $X$ and an
arbitrary polynomial tree-level superpotential $\Tr W(X)$ such that
\be\label{Wdef} W'(z) = \sum_{k=0}^{d} g_{k}z^{k} =
g_{d}\prod_{i=1}^{d}(z-w_{i})\, .\ee
The solution of this model can be generalized to many other $\nn=1$
gauge theories with various gauge groups and matter contents. The
usual approach is to use the Dijkgraaf-Vafa matrix model \cite{DV}, or
equivalently the generalized Konishi anomaly equations supplemented
with an appropriate glueball effective superpotential \cite{CDSW}.
These approaches have been motivated by some perturbative calculations
\cite{pert,CDSW}. Here perturbative is with respect to the gauge
coupling constant. Equivalently, the gauge field in \cite{pert,CDSW}
is treated as an external classical background field. This is clearly
inadequate to derive exact non-perturbative results. Our main interest
is actually in computing the expectation values of various chiral
operators, which do not have perturbative corrections!

In the present paper, we provide a non-perturbative check of the
matrix model and the anomaly equations up to the second order in the
instanton expansion. An exact proof to all orders, that applies to all
the vacua of the theory, will be presented in a forthcoming paper
\cite{mic3}. Our explicit calculations show how remarkable it is for
the anomaly equations to retain their perturbative form, at the
expense of a non-perturbative redefinition of the variables as
explained in \cite{ferchiral}. In particular, the generators of the
equations, which form perturbatively a truncated super-Virasoro
algebra, get extremely strong quantum corrections due to the
non-linearity of the associated transformations. Their action does not
close in the chiral ring, and to obtain a closed algebra one needs to
enlarge considerably the set of generators.

The full set of non-trivial expectation values in the theory
\eqref{Wdef} is given by \cite{CDSW}
\be\label{uvdef} u_{n} = \bigl\langle\Tr X^{n}\bigr\rangle\, ,\quad
v_{n} = -\frac{1}{16\pi^{2}}\bigl\langle\Tr
W^{\alpha}W_{\alpha}X^{n}\bigr\rangle\, ,\ee
where $W^{\alpha}$ is the vector chiral superfield whose lowest
component is the gluino field. It is convenient to work with the
generating functions
\be\label{genfdef} R(z;\a,q) = \sum_{n\geq 0}\frac{u_{n}}{z^{n+1}}\,
\cvp\quad S(z;\a,\g,q) = \sum_{n\geq 0}\frac{v_{n}}{z^{n+1}}\,
\cdotp\ee
We have indicated explicitly the dependence on the couplings $g_{k}$,
denoted collectively by $\g$, the instanton factor
\be\label{qdef} q = \La^{2N}\, ,\ee
and the boundary conditions at infinity for the chiral superfield $X$,
\be\label{bcdef} X_{\infty} = \diag (a_{1},\ldots,a_{N}) = \diag\a\,
.\ee
The function $R(z;\a,q)$ does not depend on $\g$ \cite{fucito} and can
be computed exactly using the results of \cite{nekrasova,nekrasovb}.
It was shown in \cite{mic1} that, on the extrema of $\wmic (\a,\g,q)$,
$R(z)$ coincides with the result obtained from the matrix model. On
the other hand, very little is known about the generalized glueball
operators $v_{n}$ for arbitrary $\a$ and $n$ (the case $n=0$ was
discussed in \cite{mic1}). The study of the generating function
$S(z;\a,\g,q)$ will thus be a central topic in the present work. An
important goal is to show that it coincides with the matrix model
prediction on-shell (i.e.\ on the extrema of $\wmic$).

The plan of the paper is as follows. In Section 2, we explain the
general set-up and introduce Nekrasov's $\Omega$-background, the
localization formulas and the sum over colored partitions that we use
to perform our calculations. We have been very careful in obtaining
the relevant equations, which can be found in the literature in many
different, and often erroneous, forms. We give general formulas for
the generating functions $R(z;\a,q)$, $S(z;\a,\g,q)$ and the
microscopic quantum superpotential $\wmic(\a,\g,q)$. In Section 3, we
present our explicit two-instanton calculations in the
$\Omega$-background. In Section 4, we focus on the anomaly equations.
After a general discussion of the non-perturbative properties of these
equations, we derive the quantum generators and algebra that generate
the equations. We show that the results are consistent with the
Dijkgraaf-Vafa matrix model and glueball superpotential. We present
our conclusions in Section 5. A technical appendix is also included at
the end of the paper.

\section{General set-up}
\setcounter{equation}{0}
\subsection{Quantum superpotential and correlators}

The microscopic quantum superpotential $\wmic(\a)$ is defined
\cite{mic1} by the following euclidean path integral with given
boundary conditions at infinity \eqref{bcdef},
\be\label{wmicdef} e^{-\int\!\d^{4}x\left( 2N\re\int\!\d^{2}\theta\,
\wmic (\a,\g,q) + D\text{-terms}\right)} = \int_{X_{\infty} = \diag\a}
\! \d\mu\, e^{-{\mathcal S}_{\text E}} \, , \ee
where ${\mathcal S}_{\text E}$ is the euclidean super Yang-Mills
action and $\d\mu$ the path integral measure including the ghosts.
It 
is shown in \cite{mic1} that
\be\label{Wmicresa}\wmic(\a,\g,q) = \vevab{\Tr W(X)}\, ,\ee
where the expectation value $\vevas{\mathscr O}$ of any
chiral operator $\mathscr O$ is defined by
\be\label{vevdef}\vevab{\mathscr O} =\frac{\int_{X_{\infty} =
\diag\a} \!
\d\mu\,\mathscr O e^{-{\mathcal S}_{\text E}}}{\int_{X_{\infty} =
\diag\a} \! \d\mu\, e^{-{\mathcal S}_{\text E}}} = \mathscr
O(\a,\g,q)\, . \ee
Equation \eqref{Wmicresa} follows from the $\u_{\text R}$ symmetry of
the theory, for which the charges of the superspace coordinates
$\theta^{\alpha}$, instanton factor $q$, chiral superfield $X$, vector
superfield $W^{\alpha}$, boundary conditions $\a$, couplings $\g$ and
superpotential $\wmic$ are given by
\be\label{charges}
\begin{matrix}
& \theta^{\alpha} & q & X & W^{\alpha} & \a & \g & \wmic \\
\uR & 1 & 0 & 0 & 1 & 0 & 2 &\hphantom{,\,} 2 \, .
\end{matrix}\ee
By varying the highest components of the chiral superfields $\g$ and
$q$ in \eqref{wmicdef}, we derive the fundamental formulas
\begin{align}\label{derwm1}n\frac{\partial\wmic}{\partial g_{n-1}}
&=\vevab{\Tr X^{n}} = u_{n}(\a,\g,q)\, ,\\ \label{derwm2}
Nq\frac{\partial\wmic}{\partial q} &=-\frac{1}{16\pi^{2}}\vevab{\Tr
W^{\alpha}W_{\alpha}} = v_{0}(\a,\g,q)\, .\end{align}
The gauge theory expectation values are obtained by going on-shell,
\be\label{qem}\frac{\partial\wmic}{\partial a_{i}} = 0\, .\ee
These equations have in general many solutions for $\a$, each
corresponding to a vacuum $|\a\rangle=|0\rangle$ of the quantum gauge
theory \cite{mic1}.

\subsection{Instantons and localization}

The expectation values $\vevas{\mathscr O}$ are analytic functions of
the variables $a_{i}$. Thus, if we can compute them in an open set in 
$\a$-space, then their values for arbitrary $\a$ can be obtained by
analytic continuation. In the region
\be\label{weakc} |a_{i}-a_{j}|\gg|\La|\ee
the theory is weakly coupled and the path integral \eqref{vevdef}
localizes on instanton configurations,
\be\label{vevinst}\mathscr O(\a,\g,q) =\frac{\sum_{k\geq
0}\int_{X_{\infty} = \diag\a} \! \d m^{(k)}\,\mathscr O (\m^{(k)})
e^{-{\mathcal S}_{\text E}}}{\sum_{k\geq 0}\int_{X_{\infty} = \diag\a}
\! \d m^{(k)}\, e^{-{\mathcal S}_{\text E}}} = \sum_{k\geq 0}\mathscr
O^{(k)}(\a,\g)\, q^{k}\, .\ee
We have denoted by $\d m^{(k)}$ the measure on the finite dimensional
moduli space of instantons of topological charge $k$ and $\mathscr
O(\m^{(k)})$ the value of the operator $\mathscr O$ for the moduli
$\m^{(k)}$. The moduli space integrals are in general ambiguous due to
small instanton singularities (see for example the first reference in
\cite{instrev}, Section VII.2). For example, the expectation values
\eqref{uvdef} are ambiguous for $n\geq 2N$. To lift these ambiguities,
we consider the non-commutative deformation of the instanton moduli
space. This yields natural definitions for the operators \eqref{uvdef}
at any $n$ \cite{ferchiral}. This crucial point will be further
discussed in Section 4. Note that while turning on the non-commutative
deformation $\vartheta\not = 0$ is necessary to define the chiral
operators at the non-perturbative level, their expectation values do
not depend on $\vartheta$ which is a real parameter.

A very important property is that the instanton series always have a
non-zero radius of convergence. This shows that $\mathscr O(\a,\g,q)$
can be obtained exactly by summing up the series in \eqref{vevinst}.
Of course, computing the moduli space integrals for any values of $k$
is a priori extremely difficult.

The calculation can be drastically simplified by using localization
techniques \cite{instc}. The idea is that the effective action for the
instantons can be written in the form
\be\label{BRS} {\mathcal S}_{\text E} = Q\cdot\Xi + \Gamma\ee
with $Q\cdot\Gamma = 0$, for some particular nilpotent linear
combination $Q$ of the supercharges. The integrals over the instanton
moduli space of $Q$-closed operators (which include the chiral
operators we are interested in) then localize on the solutions to
\be\label{fixed1} Q\cdot\Xi = 0\, .\ee
The fixed points of $Q$ can be found explicitly \cite{instc}. They
correspond to $\u$ non-commutative instantons which, in the
commutative limit $\vartheta\rightarrow 0$, go to point-like singular
instanton configurations. The remaining integrals over the moduli
space of $\u$ non-commutative instantons are simpler than the original
integrals in \eqref{vevinst}, but their explicit evaluation remains a
difficult challenge that has been solved only at topological charges
$k\leq 2$.

Very fortunately, it is possible to improve the localization
techniques by putting the theory in the so-called $\Omega$-background
\cite{nekrasova}. This background is characterized by an antisymmetric
matrix $\Omega_{\mu\nu}$ that we can choose to be of the form
\be\label{Omegamat}\Omega=\eps\begin{pmatrix} 0 & -1 & 0 & 0\\ 1 & 0 &
0 & 0\\ 0 & 0 & 0 & 1\\ 0 & 0 & -1 & 0 \end{pmatrix}\, .\ee
The complex parameter $\epsilon$ measures the strength of the
background (it is also often denoted by $\hbar$ in the literature). A
non-zero $\Omega$-background breaks Lorentz invariance and the usual
supersymmetry. For example, the standard kinetic term for the field
$X$ is replaced by
\be\label{exOmega} \Tr\bigl( D_{\mu}X - \Omega_{\nu\lambda}x_{\lambda}
F_{\mu\nu}\bigr) \bigl( D_{\mu}X^{\dagger} -
\Omega^{\dagger}_{\nu\lambda}x_{\lambda} F_{\mu\nu}\bigr)\, .\ee
However, an appropriate deformation of $Q$, that we denote by
$Q_{\eps}$, is preserved, and the action keeps the form \eqref{BRS}
with $\eps$-modified quantities. The trully remarkable fact
\cite{nekrasova} is that the solutions to the new localization problem
associated with $Q_{\eps}$ are now labeled by discrete indices. This
means that the integrals in \eqref{vevinst} are reduced to finite
sums!

\subsection{Colored partitions}

Let us describe in details the configurations that contribute
\cite{nekrasova}. First, a given topological charge $k$ can be
distributed amongst the $N$ possible $\u$ non-commutative instantons
corresponding to the $N$ $\u$ factors of the unbroken gauge group (for
arbitrary $\a$),
\be\label{part1} k = \sum_{i=1}^{N}k_{i}\, .\ee
To each integer $k_{i}\geq 0$, we associate a partition
\be\label{part2} k_{i} = \sum_{\alpha\geq 0}k_{i,\alpha}\, ,\ee
with
\be\label{part3} k_{i,1}\geq k_{i,2}\geq\cdots\geq k_{i,\tilde
k_{i,1}}> k_{i,\tilde k_{i,1}+1}=0\, .\ee
The largest integer $\alpha$ such that $k_{i,\alpha}\not = 0$ is
denoted by $\tilde k_{i,1}$, for reasons to become clear later. A
collection of integers $k_{i,\alpha}$ satisfying \eqref{part3} will be
symbolically denoted by $\mathsf k_{i}$ and the size of the partition
$\mathsf k_{i}$ is defined to be
\be\label{sizedef} |\mathsf k_{i}| = k_{i} = \sum_{\alpha=1}^{\tilde
k_{i,1}}k_{i,\alpha}\, .\ee
A colored partition $\vec{\mathsf k}$ of size
\be\label{lengcol} |\vec{\mathsf k}| = \sum_{i=1}^{N} |\mathsf
k_{i}|\ee
is a collection
\be\label{coldef} \vec{\mathsf k} = (\mathsf k_{1},\ldots,\mathsf
k_{N})\ee
of $N$ partitions $\mathsf k_{i}$. The fundamental result
\cite{nekrasova} is that \emph{the most general instanton
configurations that contribute in the topological $k$ sector can be
labeled by colored partitions of size $k=|\vec{\mathsf k}|$.}

In particular, the partition function $Z_{\eps}$ in an arbitrary
$\Omega$-background can be written as
\be\label{partfunc} Z_{\eps} =\sum_{k\geq 0}\int_{X_{\infty} =
\diag\a} \! \d m^{(k)}\, e^{-{\mathcal S}_{\text E}} = \sum_{k\geq
0}Z_{\eps}^{(k)}\, q^{k}\, ,\ee
with
\be\label{partfunc2} Z_{\eps}^{(k)} = \sum_{|\vec{\mathsf k}| = k}
\mu_{\vec{\mathsf k}}^{2}\, .\ee
The sum in \eqref{partfunc2} is over all colored partitions of size
$k$, and $\mu_{\vec{\mathsf k}}^{2}$ is a measure factor on the set of
colored partitions that we describe below. As the notation suggests,
$\mu_{\vec{\mathsf k}}^{2}$ is positive definite when $\eps$ and the
$a_{i}$s are chosen to be real. The correlators \eqref{vevinst} in an
arbitrary background are expressed in a similar way,
\be\label{corr}\vevab{\mathscr O}_{\eps}= \mathscr
O_{\eps}(\a,\g,q,\eps) = \frac{1}{Z_{\eps}}\sum_{k\geq
0}q^{k}\sum_{|\vec{\mathsf k}| = k} \mu_{\vec{\mathsf k}}^{2}\,
\mathscr O_{\vec{\mathsf k}}\, ,\ee
where $\mathscr O_{\vec{\mathsf k}}$ describes the operator $\mathscr 
O$ in the configuration $\cpart$.

\begin{figure}
\centerline{\epsfig{file=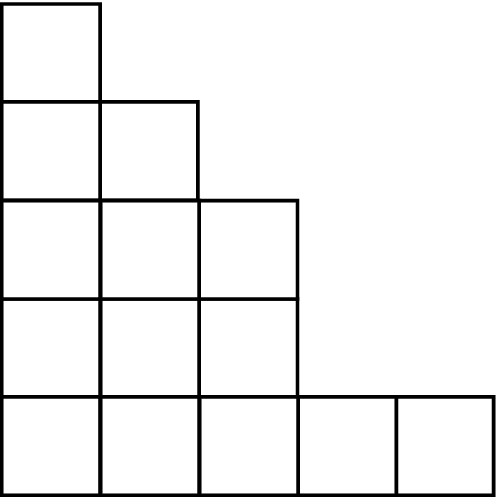,width=5cm}}
\caption{The Young tableau $Y_{\mathsf k}$ associated with the
partition $\mathsf k$ in \eqref{expart}, with integers $(k_{\alpha}) =
(5,3,3,2,1)$ and $(\tilde k_{\beta}) = (5,4,3,1,1)$.
\label{figY}}
\end{figure}

It is convenient to introduce the Young tableaux associated with the
partitions $\mathsf k_{i}$ in $\cpart$. The Young tableau associated
with any partition $\mathsf k$ is a collection of boxes arranged in
rows, the row number $\alpha$ containing $k_{\alpha}$ boxes. For
example, we have depicted in Figure \ref{figY} the Young tableau
associated with the partition
\be\label{expart} 14 = 5 + 3+ 3+2+1\, .\ee
In addition to the numbers $k_{\alpha}$ of boxes in the rows, it is
useful to also introduce the numbers $\tilde k_{\beta}$ of boxes in
the columns, with
\be\label{ktildec} \tilde k_{1}\geq \tilde
k_{2}\geq\cdots\geq \tilde k_{k_{1}} > \tilde k_{k_{1}+1}=0\, .\ee
The integers $\tilde k_{\beta}$ correspond to the number of boxes in
the rows of a partition $\tilde{\mathsf k}$ called the dual of
$\mathsf k$. Clearly
\be\label{kid} |\mathsf k| = \sum_{\alpha= 1}^{\tilde k_{1}}k_{\alpha}
= \sum_{\beta=1}^{k_{1}}\tilde k_{\beta} = |\tilde{\mathsf k}|\, .\ee
Let us now consider the box $\Box_{(\alpha,\beta)}$ in a tableau
$Y_{\mathsf k}$ belonging to the row number $\alpha$ and column number
$\beta$. The Hook length of this box is defined to be
\be\label{Hookdef} h\bigl(\Box_{(\alpha,\beta)}\bigr) = k_{\alpha} -
\beta + \tilde k_{\beta} - \alpha +1\, .\ee
Geometrically, $h(\Box)$ represents the number of boxes above and to
the right of $\Box$ in the tableau plus one.

We can now give the formula for the measure factor $\mu_{\cpart}$. Let
us start with the case $N=1$, where only ordinary partitions are
involved. Then the measure is simply given in terms of the dimension
$\dim R_{\mathsf k}$ of the irreducible representation of the
symmetric group associated with the Young tableau $Y_{\mathsf k}$,
\be\label{mukN10} \eps^{|\mathsf k|}\mu_{\mathsf
k} = \frac{1}{|\mathsf k|!}\dim R_{\mathsf k}\, .\ee
Explicitly,
\be\label{mukN1} \eps^{|\mathsf k|}\mu_{\mathsf k} =
\frac{1}{\prod_{\Box\in Y_{\mathsf k}}h(\Box)}\, \cvp\ee
where the product is taken over all the boxes in the Young tableau.
For example, for the diagram in Figure \ref{figY},
\be\label{mukex} \eps^{14}\mu_{\mathsf k} = \frac{1}{9\cdot 7\cdot
5\cdot 2\cdot 1\cdot 6\cdot 4\cdot 2\cdot 5\cdot 3\cdot 1\cdot 3\cdot
1\cdot 1} =\frac{1}{1360800}\,\cdotp \ee
It is possible to write \eqref{mukN1} is an alternative form which is 
sometimes useful,
\be\label{mukN1bis} \eps^{|\mathsf k|}\mu_{\mathsf k} =
\frac{\prod_{1\leq\alpha_{1}<\alpha_{2}\leq\tilde
k_{1}}(k_{\alpha_{1}}-k_{\alpha_{2}}-
\alpha_{1}+\alpha_{2})}{\prod_{\alpha=1}^{\tilde k_{1}}(\tilde k_{1} +
k_{\alpha} - \alpha)!}\,\cdotp\ee
The equivalence between \eqref{mukN1bis} and \eqref{mukN1} can be
shown straightforwardly by using a recursive argument on the number of
columns of the Young tableau. A generalization of this result is
proven in the Appendix. For example, in the case of Figure \ref{figY},
\eqref{mukN1bis} yields
\be\label{mukex2} \eps^{14}\mu_{\mathsf k} = \frac{3\cdot 4\cdot
6\cdot
8\cdot 1\cdot 3\cdot 5\cdot 2\cdot 4\cdot
2}{9!\,6!\,5!\,3!\,1!}=\frac{1}{1360800}\,\cvp \ee
consistently with \eqref{mukex}. 

For arbitrary $N$, the measure is given by a ``colored''
generalization of \eqref{mukN1},
\begin{multline}\label{mukgen} \mu_{\cpart} =
\prod_{i=1}^{N}\biggl[\mu_{\mathsf
k_{i}}\prod_{\Box_{(\alpha,\beta)}\in Y_{\mathsf
k_{i}}}\prod_{j\not =
i}\frac{1}{a_{i}-a_{j}+\eps(\beta-\alpha)}\biggr]\times
\\
\prod_{i<j}\prod_{\alpha=1}^{\tilde k_{i,1}}\prod_{\beta=1}^{k_{j,1}}
\frac{\bigl( a_{i}-a_{j}+\eps (\tilde k_{j,\beta}-\alpha-\beta
+1)\bigr)\bigl(a_{i}-a_{j}+\eps(k_{i,\alpha}-\beta-\alpha+1)\bigr)}
{\bigl(a_{i}-a_{j}+\eps(1-\alpha-\beta)\bigr)
\bigl(a_{i}-a_{j}+\eps(\tilde k_{j,\beta}-\alpha+ k_{i,\alpha}-\beta
+1)\bigr)}\,\cdotp\end{multline}
This formula can also be rewritten in a form analogous to
\eqref{mukN1bis},
\begin{multline}\label{mukgenbis} \mu_{\cpart}
= (-1)^{\sum_{i=1}^{N}(i-1)|\mathsf k_{i}|}\prod_{i=1}^{N}\mu_{\mathsf
k_{i}}\times \prod_{i<j}\Biggl[\prod_{\alpha_{1}=1}^{\tilde k_{i,1}}
\prod_{\alpha_{2}=1}^{\tilde k_{j,1}} \frac{
a_{i}-a_{j}+\eps(k_{i,\alpha_{1}}-k_{j,\alpha_{2}}
-\alpha_{1}+\alpha_{2})}{a_{i}-a_{j}+ \eps(\alpha_{2}-\alpha_{1})}\\
\prod_{\Box_{(\alpha,\beta)}\in Y_{\mathsf k_{i}}}
\frac{1}{a_{i}-a_{j}+\eps(\beta-\alpha + \tilde k_{j,1})}
\prod_{\Box_{(\alpha,\beta)}\in Y_{\mathsf k_{j}}}
\frac{1}{a_{i}-a_{j}-\eps(\beta-\alpha + \tilde k_{i,1})}\Biggr]\, .
\end{multline}
This form has the advantage of making $\mu_{\cpart}^{2}$ manifestly
symmetric under permutation,
\be\label{perm} a_{i}\leftrightarrow a_{j}\, ,\quad \mathsf
k_{i}\leftrightarrow\mathsf k_{j}\, ,\ee
which is a consequence of gauge invariance. It is also more convenient
to study the $\eps\rightarrow 0$ limit. The proof of the equality
between \eqref{mukgen} and \eqref{mukgenbis} is given in the Appendix.

\subsection{The scalar operators}
\label{scasec}

The operators $\Tr X^{n}$ were studied in \cite{unop} for the $\nn=2$
theory. In the configuration $\cpart$, they are given by
\begin{multline}\label{unmic} u_{n,\cpart} = \sum_{i=1}^{N}\biggl[
a_{i}^{n} + \sum_{\alpha=1}^{\tilde k_{i,1}}\Bigl(\bigl( a_{i} +
\eps(k_{i,\alpha}-\alpha+1)\bigr)^{n} - \bigl(a_{i}+\eps
(k_{i,\alpha}-\alpha)\bigr)^{n}\\ + \bigl( a_{i} -
\eps\alpha\bigr)^{n}
- \bigl( a_{i}-\eps(\alpha-1)\bigr)^{n}\Bigr)\biggr]\, .\end{multline}
It is shown in \cite{fucito}, and will be reviewed below, that this
formula remains valid in the $\nn=1$ theory as well.

The gauge theory correlators $\vevas{\Tr X^{n}}$, and thus the quantum
superpotential \eqref{Wmicresa}, can be obtained in principle from the
above formulas by taking the $\eps\rightarrow 0$ limit,
\be\label{unform} \vevab{\Tr X^{n}}=\lim_{\eps\rightarrow
0}\vevab{\Tr X^{n}}_{\eps} = \lim_{\eps\rightarrow
0}\frac{1}{Z_{\eps}}\sum_{k\geq
0}q^{k}\sum_{|\cpart|=k}\mu_{\cpart}^{2}\,u_{n,\cpart}\, .\ee
This limit was studied in \cite{nekrasovb} by using the saddle point
method. The saddle point corresponds to a very large colored
partition, of size $|\cpart|\sim 1/\eps^{2}$, for which the shapes of
the associated Young tableaux can be computed exactly. The result
\cite{nekrasovb} shows that the generating function is given by
\be\label{Rform} R(z;\a,q) = \frac{P'(z)}{\sqrt{P(z)^{2} -
4q}}\,\cdotp\ee
It is a meromorphic function on the Seiberg-Witten curve
\be\label{SWcurve} \mathcal C :\ y^{2} = P(z)^{2} - 4 q =
\prod_{i=1}^{N}(z-x_{i})^{2} - 4 q\, .\ee
This curve is a two-sheeted covering of the complex $z$-plane, with
branch cuts running from $x_{i}^{-}$ to $x_{i}^{+}$ with
\be\label{Ppmdef} P(z)\mp 2 q^{1/2} = \prod_{i=1}^{N}(z-x_{i}^{\pm})\,
.\ee
The parameters $x_{i}$ are determined in terms of the boundary
conditions $a_{j}$ by the equations
\be\label{aidef} a_{i} = \frac{1}{2i\pi}\oint_{\alpha_{i}}zR(z)\,\d
z\, ,\ee
where the closed contour $\alpha_{i}$ encircles the cut from
$x_{i}^{-}$ to $x_{i}^{+}$.

\subsection{Geometric formulation}

There is a nice geometric formulation of the localization on the
instanton moduli space that uses the notion of equivariant
differential forms. Details on this theory can be found for example in
\cite{equivbook}. We shall need only a few qualitative features, that
were also used in \cite{equiv,fucito}. The idea is that
$Q_{\eps}$-closed operators correspond to equivariantly closed forms
with respect to the symmetry transformation generated by $Q_{\eps}$.
For our purposes, the important part of this symmetry is a space-time
rotation that enters when the $\Omega$-background is turned on. It is
generated by the vector field
\be\label{symequiv} \xi = \Omega_{\mu\nu}
x_{\nu}\frac{\partial}{\partial x_{\mu}} =\eps\Bigl(
iz_{1}\frac{\partial}{\partial z_{1}}-i\bar
z_{1}\frac{\partial}{\partial
\bar z_{1}} -iz_{2}\frac{\partial}{\partial z_{2}}
+i\bar z_{2}\frac{\partial}{\partial \bar z_{2}}\Bigr)\, .\ee
The complex coordinates $z_{1}$ and $z_{2}$ are defined by
\be\label{zdef} z_{1} = x_{1} + i x_{2}\, ,\quad z_{2} = x_{3} + i
x_{4}\, .\ee
Important equivariant forms (i.e., forms that are invariant under the
transformation $z_{1}\rightarrow e^{i\gamma}z_{1}$, $z_{2}\rightarrow
e^{-i\gamma}z_{2}$ generated by $\xi$) on space-time are given
by\footnote{These forms appear in \cite{fucito}, and we have simply
corrected a minus sign.}
\begin{align}\label{ef1} \alpha_{(0,0)} & = 1\\ \label{ef2}
\alpha_{(2,0)} & = \d z_{1}\wedge\d z_{2} + i\eps z_{1}z_{2}\, ,\\
\label{ef3} \alpha_{(0,2)} & = \d\bar z_{1}\wedge\d\bar z_{2} - i\eps
\bar z_{1}\bar z_{2}\, ,\\ \label{ef4} \alpha_{(2,2)}&  = 
\d z_{1}\wedge\d z_{2}\wedge \d\bar z_{1}\wedge\d\bar z_{2} +
i\eps\bigl( z_{1}z_{2}\d\bar z_{1}\wedge\d\bar z_{2} - 
\bar z_{1}\bar z_{2}\d z_{1}\wedge\d z_{2}\bigr) + \eps^{2}
z_{1}z_{2}\bar z_{1}\bar z_{2}\, .\end{align}
It is trivial to check that all these forms are equivariantly closed,
\be\label{equivfor} \bigl(\d - i_{\xi}\bigr)\alpha_{(n,m)} = 0\,
.\ee
Equivariantly closed forms on $\mathbb C^{2}\times\mathscr M^{(k)}$,
where $\mathbb C^{2}$ is the space-time and $\mathscr M^{(k)}$ the
instanton moduli space, can then be built from the equivariant field
strength $\mathcal F$ \cite{beaulieu}. The field strength $\mathcal F$
is expressed in terms of the ADHM data; it is a linear combination of
the usual Yang-Mills field strength, gluino, scalar and fermion in the
chiral multiplet $X$ such that
\begin{align}\label{opun} &\int\!\d^{4} x\,\Tr X^{n} = \int_{\mathbb
C^{2}}\alpha_{(2,2)}\wedge\Tr\mathcal F^{n}\, ,\\ \label{opvn}
&\int\!\d^{4} x\, \Tr W^{\alpha}W_{\alpha}X^{n}
=\frac{16\pi^{2}}{(n+1)(n+2)}
\int_{\mathbb C^{2}}\alpha_{(0,2)}\wedge\Tr\mathcal F^{n+2}\, ,\\
\label{opunW} &\int\!\d^{4} x\int\!\d^{2}\theta\,
\Tr X^{n} =\int_{\mathbb C^{2}}\alpha_{(2,0)}\wedge\Tr\mathcal F^{n}\,
,\\ \label{opvnW} &\int\!\d^{4} x\int\!\d^{2}\theta\, \Tr
W^{\alpha}W_{\alpha}X^{n} =\frac{16\pi^{2}}{(n+1)(n+2)} \int_{\mathbb
C^{2}}\Tr\mathcal F^{n+2}\, .\end{align}

Equations \eqref{opun}, \eqref{opvn} and \eqref{opunW} were obtained
in \cite{fucito} (we have simply put the correct factors to match with
our conventions). Equation \eqref{opvnW} can be obtained similarly by
a straightforward calculation from the explicit expression for
$\mathcal F$.

The integral of an equivariantly closed form localizes on the fixed
point of the associated symmetry transformation \cite{equivbook}. All
we need is that, for any form $\alpha$ satisfying \eqref{equivfor},
\be\label{locform} \int_{\mathbb C^{2}}\alpha =
\frac{1}{\eps^{2}}\alpha^{(0)}\, ,\ee
where $\alpha^{(0)}$ is the zero-form part of $\alpha$ evaluated at
the origin $O$ of space-time where the vector \eqref{symequiv}
vanishes.\footnote{We define the integral $\int_{\mathbb C^{2}}$ in
such a way that there is no overall constant factor in
\eqref{locform}.} One must be careful in applying this rule because we
have regulated the integrals over the instanton moduli space by
formulating the theory on a non-commutative space-time. The
coordinates $z_{1}$ and $z_{2}$ are really operators satisfying
\be\label{NCzdef} \bigl[\hat z_{a},\hat{\bar z}_{b}\bigr] =
\vartheta\delta_{ab}\, ,\ee
for which we can use the representation
\be\label{NCzfor} \hat z_{a} = \vartheta\frac{\partial}{\partial\bar
z_{a}}\,\cdotp\ee
For example, if we compute the volume of space-time using the form
\eqref{ef4} and \eqref{locform}, we find
\be\label{stvol} V = \int_{\mathbb C^{2}}\d z_{1}\wedge\d z_{2}\wedge
\d\bar z_{1}\wedge\d\bar z_{2} = \int_{\mathbb C^{2}}\alpha_{(2,2)}
=\frac{1}{\eps^{2}}\eps^{2}\hat z_{1}\hat z_{2}\hat{\bar
z}_{1}\hat{\bar z}_{2}|_{O} = \vartheta^{2}\, .\ee
The same calculation for the integral in the right hand side of
\eqref{opun} yields
\be\label{exequi} \int\!\d^{4} x\,\Tr X^{n} =
\vartheta^{2}\bigl(\Tr\mathcal F^{n}\bigr)^{(0)} = \vartheta^{2}\Tr
X^{n}\, ,\ee
showing that the zero-form part of $\Tr\mathcal F^{n}$ is given by
\be\label{Fzero} \bigl(\Tr\mathcal F^{n}\bigr)^{(0)} = \Tr X^{n}\,
.\ee
This result will be useful later.

Another simple application is to derive the result of \cite{fucito}
that the $\vevas{\Tr X^{n}}_{\eps}$ do not depend on $\g$. We write
the euclidean action as a sum of two terms, the $\nn=2$ action that
does not depend on the couplings $\g$ and the $\nn=1$ superpotential
term,
\be\label{Seucldec}\begin{split} \mathcal S_{\text E} & = \mathcal
S_{\nn=2} + N\int\!\d^{4}x\int\!\d^{2}\theta \Tr W(X) + \text{c.\
c.}\\ & = \mathcal S_{\nn=2} + N\int_{\mathbb
C^{2}}\alpha_{(2,0)}\wedge\Tr W(\mathcal F) + \text{c.\
c.}\end{split}\ee
The overall factor of $N$ in \eqref{Seucldec} is a natural convention,
consistent with \eqref{wmicdef} and \eqref{Wmicresa}, that makes the
action of order $N^{2}$. We have also used \eqref{opunW} to rewrite
the superpotential term as the integral of an equivariantly closed
form. We shall no longer indicate explicitly the anti-chiral terms in
the following (the $+\text{c.\ c.}$ in \eqref{Seucldec}), since they
obviously do not contribute to the chiral operators expectation
values. The idea is now to expand the factor $e^{-\mathcal S_{\text
E}}$ in the path integral in powers of $W$ and then to apply the
localization formula \eqref{locform}. Since the zero-form part of
$\alpha_{(2,0)}$ contains only $z_{1}z_{2}$, a $p^{\text{th}}$ power
of $W$ yields $(z_{1}z_{2})^{p}$. On the other hand, the insertion of
$\Tr X^{n}$ yields, according to \eqref{opun} and \eqref{ef4}, a
factor of $z_{1}z_{2}\bar z_{1}\bar z_{2}$. Taking into account the
non-commutativity, we have to compute
\be\label{NCcalc1} \bigl(\hat z_{1}\hat z_{2}\bigr)^{p+1}\hat{\bar
z}_{1}\hat{\bar z}_{2}|_{O} = \vartheta^{2}\delta_{p,0}\, ,\ee
showing that there is no dependence in $W$. The same reasoning also
shows that the correlators $\vevas{\Tr X^{n_{1}}\cdots \Tr
X^{n_{s}}}_{\eps}$ are independent of $\g$ as well. This is
non-trivial because the multi-trace correlators do not factorize at
finite $\eps$ but only in the $\eps\rightarrow 0$ limit.

\subsection{The glueball operators}
\label{glueop}

Let us now derive the basic formula for the expectation values of the
generalized glueball operators,
\begin{multline}\label{vnform} -\frac{1}{16\pi^{2}}\vevab{\Tr
W^{\alpha}W_{\alpha}X^{n}}_{\eps} =
\frac{N}{(n+1)(n+2)}\frac{1}{\eps^{2}}\Bigl( \vevab{\Tr W(X) \Tr
X^{n+2}}_{\eps}\\ - \vevab{\Tr W(X)}_{\eps}\vevab{\Tr
X^{n+2}}_{\eps}\Bigr)\, .\end{multline}
This formula relates the glueballs to the $\vevas{\Tr X^{n}}_{\eps}$
computed in \ref{scasec}. It appears in the special case of
$W(X) = \frac{1}{2}mX^{2}$ in \cite{fucito}. Of course, we are mainly
interested in the $\eps\rightarrow 0$ gauge theory limit
\be\label{vngauge} v_{n}(\a,\g,q)=-\frac{1}{16\pi^{2}}
\lim_{\eps\rightarrow 0}\vevab{\Tr W^{\alpha}W_{\alpha}X^{n}}_{\eps}\,
.\ee
A very interesting aspect of \eqref{vnform} is to show that the
glueball expectation values are related to the \emph{subleading} terms
in the small $\eps$ expansion of $\vevas{\Tr X^{n}}_{\eps}$. This
means that the first corrections in the $\Omega$-background are
relevant to the $\nn=1$ gauge theory. In particular, the leading
$\eps\rightarrow 0$ approximation studied in \cite{nekrasovb} to solve
the $\nn=2$ theory is not sufficient for the case of $\nn=1$.

Equation \eqref{vnform} is the main starting point for the
calculations performed in Sections 3 and 4. We are going to give two
derivations. The first one follows closely the reasoning in
\cite{fucito}. The second one uses the properties of the quantum
superpotential $\wmic$. A third derivation, which is less formal and
completely explicit, will also be given in \cite{mic3} using an
extended version of the theory.

\subsubsection{First derivation}

Using \eqref{opvn}, we have
\begin{align} 
-\frac{1}{16\pi^{2}}\vevab{\Tr W^{\alpha}W_{\alpha}X^{n}}_{\eps} & =
-\frac{1}{16\pi^{2}}\int\! \frac{\d^{4} x}{V}\vevab{\Tr
W^{\alpha}W_{\alpha}X^{n}}_{\eps} \\\label{vnstart} & =
-\frac{1}{(n+1)(n+2)}\,\frac{1}{\vartheta^{2}}\,\vevab{\int_{\mathbb
C^{2}}\alpha_{(0,2)}\wedge\Tr\mathcal F^{n+2}}_{\eps}\, .
\end{align}
The zero-form part of $\alpha_{(0,2)}$ in \eqref{ef3} is proportional
to $\bar z_{1}\bar z_{2}$. From \eqref{NCcalc1}, we know that the
localization procedure can yield non-zero contributions only if this
term is saturated with another contribution in $z_{1}z_{2}$. According
to \eqref{Seucldec} and \eqref{ef2}, such a contribution can come only
from a term linear in the superpotential $W$. This is produced by
expanding $e^{-\mathcal S_{\text E}}$ to linear order in $W$. Using
\eqref{locform} and \eqref{Fzero}, we see that the numerator of
\eqref{vevinst} yields a term
\begin{multline}\label{term1}
-\frac{1}{(n+1)(n+2)}\frac{1}{\vartheta^{2}}\vevab{ -N\int_{\mathbb
C^{2}}\alpha_{(2,0)}\wedge\Tr W(\mathcal F) \int_{\mathbb
C^{2}}\alpha_{(0,2)}\wedge\Tr\mathcal F^{n+2}}_{\eps} =\\
\frac{N}{(n+1)(n+2)}\frac{1}{\vartheta^{2}}\frac{i\eps}{\eps^{2}}
\frac{-i\eps}{\eps^{2}}\vartheta^{2}\vevab{\Tr W(X)\Tr X^{n+2}}_{\eps}
=\\
\frac{N}{(n+1)(n+2)}\frac{1}{\eps^{2}}\vevab{\Tr W(X)\Tr
X^{n+2}}_{\eps}
\end{multline}
and the denominator of \eqref{Seucldec} yields
\begin{multline}\label{term2}
-\frac{1}{(n+1)(n+2)}\frac{1}{\vartheta^{2}}
\vevab{N\int_{\mathbb C^{2}}\alpha_{(2,0)}\wedge\Tr W(\mathcal
F)}_{\eps}\vevab{\int_{\mathbb C^{2}}\alpha_{(0,2)}\wedge\Tr\mathcal
F^{n+2}}_{\eps}=\\
-\frac{N}{(n+1)(n+2)}\frac{1}{\vartheta^{2}}
\frac{i\eps}{\eps^{2}}\vevab{\Tr W(X)}_{\eps}\frac{-i\eps}{\eps^{2}}
\vartheta^{2}\vevab{\Tr X^{n+2}}_{\eps} =\\
-\frac{N}{(n+1)(n+2)}\frac{1}{\eps^{2}}\vevab{\Tr
W(X)}_{\eps}\vevab{\Tr X^{n+2}}_{\eps}
\end{multline}
Combining \eqref{term1} and \eqref{term2} together, we obtain
\eqref{vnform}.

\subsubsection{Second derivation}

Let us perturb the theory by adding to the tree-level superpotential
$\Tr W(X)$ a term $-\frac{t}{16\pi^{2}}\Tr W^{\alpha}W_{\alpha}X^{n}$.
According to \eqref{opvnW}, the new euclidean action is thus
\be\label{Seuclext} \mathcal S_{\text E}= \mathcal S_{\nn=2} +
N\int_{\mathbb C^{2}}\alpha_{(2,0)}\wedge\Tr W(\mathcal F) -
\frac{Nt}{(n+1)(n+2)}\int_{\mathbb C^{2}} \Tr \mathcal F^{n+2} +
\text{c.\ c.}\ee
The formula \eqref{Wmicresa} for the quantum superpotential is still
valid for non-zero $t$ and $\eps$. This follows from the fact that $t$
and $\eps$ have charge zero under the $\u_{\text R}$ symmetry
\eqref{charges}. Moreover, we have, similarly to \eqref{derwm1} and
\eqref{derwm2},
\be\label{derwm3} \frac{\partial\wmic}{\partial t} =
-\frac{1}{16\pi^{2}}\vevab{\Tr W^{\alpha}W_{\alpha}X^{n}}_{\eps} =
\frac{\partial\vevab{\Tr W(X)}_{\eps}}{\partial t}\,\cdotp\ee
Using \eqref{opun}, this is equivalent to
\be\label{vnstart2}-\frac{1}{16\pi^{2}}\vevab{\Tr
W^{\alpha}W_{\alpha}X^{n}}_{\eps} =
\frac{1}{\vartheta^{2}}\frac{\partial}{\partial t}\vevab{\int_{\mathbb
C^{2}}\alpha_{(2,2)}\wedge\Tr W(\mathcal F)}_{\eps}\, .\ee
This identity is the starting point of our second derivation of
\eqref{vnform} (compare with the starting point \eqref{vnstart} of the
first derivation). The use of the localization procedure is
particularly simple here, because the zero-form part of
$\alpha_{(2,2)}$ is proportional to $z_{1}z_{2}\bar z_{1}\bar z_{2}$
and thus non-zero contributions can only come from terms proportional
to the trivial form \eqref{ef1}, i.e.\ from the term proportional to
$t$ in \eqref{Seuclext}. The expectation value in \eqref{vnstart2} is
given by the general formula \eqref{vevinst}. Taking the derivative of
the numerator with respect to $t$ and using \eqref{Seuclext} then
yields
\begin{multline}\label{term1bis} \frac{1}{\vartheta^{2}}
\vevab{\int_{\mathbb C^{2}}\alpha_{(2,2)}\wedge\Tr W(\mathcal
F) \frac{N}{(n+1)(n+2)}\int_{\mathbb C^{2}}\Tr\mathcal F^{n+2}}_{\eps}
=\\
\frac{N}{(n+1)(n+2)}\frac{1}{\vartheta^{2}}\frac{\eps^{2}}{\eps^{2}}
\frac{1}{\eps^{2}} \vartheta^{2}\vevab{\Tr W(X)\Tr X^{n+2}}_{\eps} =\\
\frac{N}{(n+1)(n+2)}\frac{1}{\eps^{2}}\vevab{\Tr W(X)\Tr
X^{n+2}}_{\eps}\, ,
\end{multline}
whereas the variation of the denominator yields
\begin{multline}\label{term2bis} -\frac{1}{\vartheta^{2}}
\vevab{\int_{\mathbb C^{2}}\alpha_{(2,2)}\wedge\Tr W(\mathcal F)}
\frac{N}{(n+1)(n+2)}\vevab{\int_{\mathbb C^{2}}\Tr\mathcal
F^{n+2}}_{\eps} =\\
-\frac{N}{(n+1)(n+2)}\frac{1}{\vartheta^{2}}\frac{\eps^{2}}{\eps^{2}}
\vartheta^{2}\vevab{\Tr W(X)}_{\eps}\frac{1}{\eps^{2}}\vevab{\Tr
X^{n+2}}_{\eps} =\\
-\frac{N}{(n+1)(n+2)}\frac{1}{\eps^{2}}\vevab{\Tr W(X)}_{\eps}
\vevab{\Tr X^{n+2}}_{\eps}\, .
\end{multline}
Combining \eqref{term1bis} and \eqref{term2bis}, we obtain again
\eqref{vnform} (which is valid for any value of $t$, even though we
are focusing on the $t=0$ theory).

\section{Two instanton calculations at order $\eps^{2}$}
\setcounter{equation}{0}
\subsection{The expectation values $\vevas{\Tr X^{n}\Tr
X^{m}}_{\eps}$}\label{2instexpl}

In this Section, we compute explicitly the correlators $\vevas{\Tr
X^{n}\Tr X^{m}}_{\eps}$ up to two instantons,
\be\label{2inst} u_{n,m}(\a,q,\eps) =\vevab{\Tr X^{n}\Tr X^{m}}_{\eps}
= u_{n,m}^{(0)}(\a) + u_{n,m}^{(1)}(\a,\eps)\, q +
u_{n,m}^{(2)}(\a,\eps) \,q^{2} + \mathcal O(q^{3})\, .\ee
Our main goal is to use the resulting formulas to compute the glueball
operators (Section \ref{gSec}) and to check the anomaly equations
(Section \ref{aSec}). For this purpose, we are particularly interested
in the first corrections at small $\eps$,
\be\label{epsexp} u_{n,m}^{(k)}(\a,\eps)  =
u_{n,m}^{(k,0)}(\a) +u_{n,m}^{(k,2)}(\a)\,\eps^{2} + \mathcal
O(\eps^{4})\, .\ee
Note that the functions $u_{n,m}^{(k)}(\a,\eps)$ are even in $\eps$,
to any order. This result is proven in the Appendix. Our starting
formula, which is a special case of \eqref{corr}, is given by
\be\label{genfor} u_{n,m}(\a,q,\eps) = \frac{1}{Z_{\eps}}\sum_{k\geq
0}q^{k}\sum_{|\vec{\mathsf k}| = k} \mu_{\vec{\mathsf k}}^{2}\,
u_{n,\vec{\mathsf k}}\, u_{m,\vec{\mathsf k}}\, .\ee
The various ingredients entering into this formula are defined in
\eqref{partfunc}, \eqref{partfunc2}, \eqref{mukgen} and \eqref{unmic}.
Expanding at small $q$ both the numerator and the denominator in
\eqref{genfor}, we find that
\begin{align}\label{order0} &u_{n,m}^{(0)} =u_n^{\cl} \, u_m^{\cl}
\, , \\
\label{order1} & u_{n,m}^{(1)}  =
\sum_{|\vec{\mathsf k}| = 1}
\mu_{\vec{\mathsf k}}^{2}\Bigl(u_{n}^{\cl} \bigl(u_{m,\vec{\mathsf
k}}-u_m^{\cl}\bigr) + u_m^{\cl} \bigl(u_{n,\vec{\mathsf
k}}-u_n^{\cl}\bigr) + \bigl(u_{n,\vec{\mathsf k}}-u_n^{\cl}\bigr)
\bigl(u_{m,\vec{\mathsf k}}-u_m^{\cl}\bigr)\Bigr) \, ,\\
\label{order2}&\begin{aligned}
u_{n,m}^{(2)}= \sum_{|\vec{\mathsf k}| = 2}
\mu_{\vec{\mathsf k}}^{2}\Bigl(u_n^{\cl} \bigl(u_{m,\vec{\mathsf
k}}-u_m^{\cl}\bigr) + u_m^{\cl}\bigl(u_{n,\vec{\mathsf
k}}-u_n^{\cl}\bigr)  + \bigl(u_{n,\vec{\mathsf k}}-u_n^{\cl}\bigr)
\bigl(u_{m,\vec{\mathsf k}}-u_m^{\cl}\bigr)\Bigr)\\-Z_{\eps}^{(1)}\,
u_{n,m}^{(1)}\, ,\end{aligned}\end{align}
where we have defined 
\be\label{ucl}u_n^{\cl}=\sum_{i=1}^N a_i^n\, .\ee

\noindent\textsc{One instanton}: There are $N$ colored partitions
$\cpart^{(i)}$ of size $|\cpart^{(i)}| = 1$, which describe one
instanton in each $\u$ factor of the unbroken gauge group, each
contributing one term in the sum \eqref{order1}. Explicitly,
\be\label{pof1}k_{j,\alpha}^{(i)}=\delta_{i,j}\, \delta_{\alpha,1}\,
,\quad 1\leq i\leq N\, ,\ee
and \eqref{mukgen} or \eqref{mukgenbis} then yields
\be\label{mukione}\mu_{\cpart^{(i)}}^{2} =
\frac{1}{\eps^{2}}\frac{1}{\prod_{j\not = i}(a_{j}-a_{i})^{2}}\, \cdotp\ee
From \eqref{unmic} we also get
\be\label{unkex1}u_{n,\cpart^{(i)}}= u_n^{\cl} +
\frac{n!}{(n-2)!}a_i^{n-2}\,\eps^2  + \frac{n!}{(n-4)!}
\frac{a_i^{n-4}}{12}\,\eps^4 +\frac{n!}{(n-6)!}\frac{
a_i^{n-6}}{360}\,\eps^6 + \mathcal O(\eps^8)\, .\ee
To express the result, it is convenient to introduce the notation
\be\label{aijdef} a_{ij} = a_{i}-a_{j}\, .\ee
Combining \eqref{mukione} and \eqref{unkex1} in \eqref{order1} then
yields
\begin{align}\label{oneins0}&u_{n,m}^{(1,0)} = \sum_{i}
\frac{1}{\prod_{j\neq i} a_{ij}^2} \bigg( \frac{m!}{(m-2)!}\,
u_n^{\cl} a_i^{m-2} + \frac{n!}{(n-2)!} \, u_m^{\cl}a_i^{n-2} \bigg)
\, , \\
\label{oneins2} &\begin{aligned} u_{n,m}^{(1,2)} =\sum_{i}
\frac{1}{\prod_{j\neq i} a_{ij}^2} \bigg(  
\frac{m!}{12(m-4)!}\, u_n^{\cl} a_i^{m-4} &+
\frac{n!}{12(n-4)!}\, u_m^{\cl} a_i^{n-4} \\ &+
\frac{n!m!}{(n-2)!(m-2)!}\, a_i^{n+m-4} \bigg) \, , \end{aligned}\\ 
\label{oneins4}&\begin{aligned} u_{n,m}^{(1,4)}=  \sum_{i}&
\frac{1}{\prod_{j\neq i} a_{ij}^2} \bigg(  
\frac{m!}{360(m-6)!}\, u_n^{\cl} \, a_i^{m-6} 
+\frac{n!}{360(n-6)!}\, u_m^{\cl} \, a_i^{n-6}\\&
+\frac{n!m!}{12(n-4)!(m-2)!}\,
a_i^{n+m-6}+\frac{n!m!}{12(n-2)!(m-4)!}\, a_i^{n+m-6}
\bigg)\, .\end{aligned} \end{align}
Let us note that the term $u_{n,m}^{(1,4)}$, that contributes for one
instanton at order $\eps^{4}$, also contributes at two instantons at
order $\eps^{2}$, and thus will be crucial to get the correct
two-instantons correction to the glueball operators. This $\eps^{2}$
contribution comes from the last term in \eqref{order2}, taking into
account the fact that $Z_{\eps}^{(1)} \propto 1/\eps^2$. This is a
general feature of these expansions: to get the $\eps^{2q}$ terms at
$k$-instantons, one needs to compute to order $\eps^{2(q+k-k')}$ at
$k'<k$ instantons, because $Z_{\eps}^{(k')}\propto
1/\eps^{2k'}$.\smallskip\\
\textsc{Two instantons}: The sum in \eqref{order2} has $N(N+3)/2$
terms, given by the colored partitions $\cpart^{(i)}$ and
$\cpart^{(ij)}$ characterized by
\begin{align}
\label{pof22} k^{(i)}_{j,\alpha}& = \delta_{i,j} (
\delta_{\alpha,1}+\delta_{\alpha,2})\, ,\quad 1\leq i\leq N\, ,\\
\label{pof23} k^{(ij)}_{l,\alpha} &= (\delta_{i,l}+\delta_{j,l})
\delta_{\alpha,1}\, ,\quad 1\leq i\leq j\leq N\, .\end{align}
Computing carefully $\mu_{\cpart^{(i)}}^{2}$,
$\mu_{\cpart^{(ij)}}^{2}$, $u_{n,\cpart^{(i)}}$ and
$u_{n,\cpart^{(ij)}}$ from \eqref{mukgen} and \eqref{unmic}, and
plugging into \eqref{order2}, we find the following explicit
two-instantons result at order $\eps^{2}$,
\be\label{twoinst0}\begin{aligned}u_{n,m}^{(2,0)}& = 
m(m-1)\, u_n^{\cl} \Biggl[ \sum_{i} \frac{1}{\prod_{l\neq i} a_{il}^4}
\biggl( 2 \Bigl(\sum_{l\neq i} \frac{1}{a_{il}}\Big)^2a_i^{m-2}+
\sum_{l\neq i} \frac{1}{a_{il}^2}\, a_i^{m-2}\\ &\!\!-(m-2)
\sum_{l\neq i}
\frac{1}{a_{il}} \, a_i^{m-3}+\frac{(m-2)(m-3)}{4}\,
a_i^{m-4}\biggr)+\sum_{i\neq j}\frac{1}{\prod_{l\neq i}a_{il}^2\,
\prod_{l\neq j}a_{jl}^2}\frac{2a_i^{m-2}}{a_{ij}^2}\Biggr]\\ &\!\! +
\bigl( n \leftrightarrow m \bigr) + n(n-1)m(m-1)\,
\sum_{i,j}\frac{a_i^{n-2}a_j^{m-2}}{\prod_{l\neq i}a_{il}^2
\prod_{l\neq j}a_{jl}^2} \, \cvp \end{aligned}\ee
\begin{multline}\label{twoinst2}
u_{n,m}^{(2,2)} = m(m-1)\, u_n^{\cl} \Biggl[\sum_i
\frac{1}{\prod_{l\neq i}a_{il}^4}\Biggl( \biggl[
\frac{2}{3}\Bigl(\sum_{l\neq i}
\frac{1}{a_{il}}\Bigr)^4+2\Bigl(\sum_{l\neq i}
\frac{1}{a_{il}}\Bigr)^2
\sum_{l\neq i} \frac{1}{a_{il}^2}\\ + \frac{4}{3}\Bigl(\sum_{l\neq i}
\frac{1}{a_{il}}\Bigr)\sum_{l\neq i}
\frac{1}{a_{il}^3}+\frac{1}{2}\Bigl(\sum_{l\neq i}
\frac{1}{a_{il}^2}\Bigr)^2+\frac{1}{2} \sum_{l\neq i}
\frac{1}{a_{il}^4} \biggr] a_i^{m-2}\\
-\frac{m-2}{2}\biggl[\frac{4}{3}\Bigl(\sum_{l\neq i}
\frac{1}{a_{il}}\Bigr)^3+2\Bigl(\sum_{l\neq i}
\frac{1}{a_{il}}\Bigr)\sum_{l\neq i}
\frac{1}{a_{il}^2}+\frac{2}{3}\sum_{l\neq i} \frac{1}{a_{il}^3}\biggr]
a_i^{m-3}\\+\frac{(m-2)(m-3)}{3}\biggl[2\Bigl(\sum_{l\neq i}
\frac{1}{a_{il}}\Bigr)^2+\sum_{l\neq i} \frac{1}{a_{il}^2}\biggr]
a_i^{m-4}-\frac{(m-2)!}{4(m-5)!}\Bigl(\sum_{l\neq i}
\frac{1}{a_{il}}\Bigr)\, a_i^{m-5}\\
+\frac{(m-2)!}{24(m-6)!}\, a_i^{m-6} \Biggr) +\sum_{i\neq
j}\frac{1}{\prod_{l\neq i}a_{il}^2 \prod_{l\neq
j}a_{jl}^2}\Biggl(\frac{3\,
a_i^{m-2}}{a_{ij}^4}+\frac{(m-2)(m-3)}{6}\frac{a_i^{m-4}}{a_{ij}^2}\Biggr)
\Biggr]+\displaybreak
\\ + \bigl( n \leftrightarrow m\bigr) 
+ n(n-1)m(m-1)\,\Biggl[ \sum_i \frac{1}{\prod_{l\neq
i}a_{il}^4}\Biggl(\frac{(n-2)(m-2)}{2}\, a_i^{n+m-6}\\
+\frac{7}{12}\Bigl((n-2)(n-3)+(m-2)(m-3)\Bigr)\, a_i^{n+m-6}
-2(n+m-4)\Bigl(\sum_{l\neq i} \frac{1}{a_{il}}\Bigr)\, a_i^{n+m-5}\\
\hskip 6cm +4\Bigl(\sum_{l\neq i} \frac{1}{a_{il}}\Big)^2\,
a_i^{n+m-4}+ 2 \Bigl(\sum_{l\neq i} \frac{1}{a_{il}^2}\Big)\,
a_i^{n+m-4} \Biggr)\\ + \sum_{i\neq j}\frac{1}{\prod_{l\neq
i}a_{il}^2 \prod_{l\neq j}a_{jl}^2}\Biggl(\frac{(n-2)(n-3)}{12}\,
a_i^{n-4}a_j^{m-2} + \frac{(m-2)(m-3)}{12}\, a_i^{m-4}\, a_j^{n-2} \\
\hskip 8cm
+\frac{2}{a_{ij}^2}\Big(a_i^{n+m-4}+a_i^{n-2}a_j^{m-2}\Bigr)
\Biggr)\Biggr]
\,.
\end{multline}
Let us note that as a special case of the above calculation, we also
find the expectation values of $\vevas{\Tr X^{n}}_{\eps}$,
\be\label{u0m} u_{n}(\a,q,\eps) = \frac{u_{n,0}(\a,q,\eps)}{N} =
u_{n}^{\cl}(\a) + u_{n}^{(1)}(\a,\eps)\, q + u_{n}^{(2)}(\a,\eps)\,
q^{2} + \mathcal O(q^{3})\, ,\ee
and in particular the microscopic quantum superpotential
\eqref{Wmicresa} is known up to two instantons.

\subsection{The glueball operators expectation values}
\label{gSec}

We can now use the fundamental formula \eqref{vnform} to get the
glueball operators expectation values, at $\eps=0$, from the results
of the previous subsection. Expanding
\be\label{2inst3}v_m(\a,\g,q) = v_m^{(1)}(\a,\g)\,
q+v_m^{(2)}(\a,\g)\, q^2+\mathcal O(q^3)\, ,  \ee
we find
\begin{align}\label{vorder1}&\begin{aligned}v_m^{(1)}(\a,\g)=&
\frac{N}{(m+1)(m+2)}\lim_{\eps \rightarrow
0}\frac{1}{\eps^2}\sum_{n\geq 0} \Bigg[ \frac{g_n}{n+1} \bigg(
u_{n+1,m+2}^{(1)}(\a,\eps)\\ &\hskip 3cm - u_{n+1}^{\cl}(\a)
u_{m+2}^{(1)}(\a,\eps) - u_{m+2}^{\cl}(\a)
u_{n+1}^{(1)}(\a,\eps)\bigg)\Bigg] \, , \end{aligned} \\
\label{vorder2}&\begin{aligned}v_m^{(2)}(\a,&\g)=
\frac{N}{(m+1)(m+2)}\lim_{\eps \rightarrow
0}\frac{1}{\eps^2}\sum_{n\geq 0}\Bigg[ \frac{g_n}{n+1} \bigg(
u_{n+1,m+2}^{(2)}(\a,\eps) \\ &- u_{n+1}^{\cl}(\a)
u_{m+2}^{(2)}(\a,\eps) - u_{m+2}^{\cl}(\a) u_{n+1}^{(2)}(\a,\eps)
-u_{n+1}^{(1)}(\a,\eps)u_{m+2}^{(1)}(\a,\eps)\bigg)\Bigg]\, .
\end{aligned} \end{align}
A careful calculation then yields the following explicit formulas, for
the one-instanton contribution,
\be\label{vm1}v_m^{(1)}(\a,\g)=N \sum_i \frac{W''(a_i)\,
a_i^{m}}{\prod_{l\neq i}a_{il}^2}\ee
and for the two-instantons contribution,
\be\label{vm2}\begin{split}v_m^{(2)}(\a,\g)=N \Bigg[ & \sum_i
\frac{1}{\prod_{l\neq i}a_{il}^4} \Bigg( \bigg[
\frac{1}{2}W''''(a_i)-2 \sum_{l\neq i}
\frac{1}{a_{il}}W'''(a_i)+4\Big(\sum_{l\neq i}
\frac{1}{a_{il}}\Big)^2 W''(a_i) \\ & +2 \sum_{l\neq i}
\frac{1}{a_{il}^2}W''(a_i)\bigg]\, a_i^m + m
\bigg[\frac{1}{2}W'''(a_i)-2\sum_{l\neq i}
\frac{1}{a_{il}}W''(a_i)\bigg]\, a_i^{m-1}\\ &+\frac{m(m-1)}{2}\,
W''(a_i)\, a_i^{m-2}\Bigg)+\sum_{i\neq
j}\frac{W''(a_i)+W''(a_j)}{\prod_{l\neq i}a_{il}^2 \prod_{l\neq
j}a_{jl}^2}\, \frac{2a_i^m}{a_{ij}^2}\Bigg]\, . \end{split}\ee

We now have all the necessary ingredients to perform the check of the
Dijkgraaf-Vafa matrix model from our purely microscopic point of view.
In principle, all we have to do is to show that the above correlators
satisfy the generalized Konishi anomaly equations when we go on-shell,
i.e.\ when we extremize $\wmic$ (of course the correlators will
\emph{not} satisfy the anomaly equations for arbitrary values of
$\a$). We are going to perform this check in the next Section, and
also exhibit highly non-trivial features of the anomaly equations at
the non-perturbative level.

\section{Non-perturbative anomaly equations}
\label{aSec}
\setcounter{equation}{0}
\subsection{Introduction}

A cornerstone of our understanding of $\nn=1$ gauge theories, and
their relation with the Dijkgraaf-Vafa matrix model, is the set of
generalized anomaly equations studied in \cite{CDSW}. These equations 
have been derived in perturbation theory (i.e.\ in a fixed classical
background gauge field) in the following way \cite{CDSW}.

We consider some particular non-linear variations of the field $X$ in
the path integral \cite{CDSW}, which are generated by the
operators
\be\label{varop} L_{n} = -X^{n+1}\frac{\delta}{\delta X}\,\cvp\quad
J_{n} = \frac{1}{16\pi^{2}} W^{\alpha}W_{\alpha}X^{n+1}
\frac{\delta}{\delta X}\,\cvp\quad\text{for}\ n\geq -1\, .\ee
In \cite{CDSW} the operators $W^{\alpha}X^{n+1}\delta/\delta X$ were
also considered, but the resulting equations do not produce
non-trivial constraints on expectation values.\footnote{We could
include them straightforwardly in the discussion by introducing
Lorentz-violating couplings $t_{n}^{\alpha}\Tr W_{\alpha} X^{n+1}$ in
the tree-level superpotential.} The operators act on the gauge
invariant observables as
\be\label{LJact} L_{n}\cdot u_{m} = -m u_{n+m}\, ,\quad
J_{n}\cdot u_{m} = - m v_{n+m}\, ,\quad  L_{n}\cdot v_{m} =
-m v_{n+m}\, ,\quad J_{n}\cdot v_{m} = 0\, ,\ee
and satisfy the algebra
\be\label{pertalg} [L_{n},L_{m}] = (n-m) L_{n+m}\, ,\quad
[L_{n},J_{m}] = (n-m) J_{n+m}\, ,\quad [J_{n},J_{m}]=0\, .\ee
The relations $J_{n}\cdot v_{m} = 0$ and $[J_{n},J_{m}]=0$ follow from
the fact that the $W^{\alpha}$ anticommutes in the chiral ring. The
anomaly polynomials generated by $L_{n}$ and $J_{n}$ are respectively 
\cite{CDSW}
\begin{align}\label{ano1}\mathscr A_{n} & = - N\sum_{k\geq 0}
g_{k}u_{n+k+1} + 2\sum_{k_{1}+k_{2} = n} u_{k_{1}}v_{k_{2}}\, , \\
\label{ano2} \mathscr B_{n} & = - N\sum_{k\geq 0} g_{k}v_{n+k+1} +
\sum_{k_{1}+k_{2} = n} v_{k_{1}}v_{k_{2}}\, .
\end{align}
The terms linear in the fields in \eqref{ano1} and \eqref{ano2} come
from the tree-level action, whereas the quadratic terms are generated
by an anomalous jacobian in the path integral measure (in the Fujikawa
approach) or equivalently by a one-loop calculation with external
gauge fields. It is not difficult to show that this result is exact in
perturbation theory, to any loop order, for example by using the
Wess-Zumino consistency conditions
\begin{align}\label{WZcc1} L_{n}\cdot \mathscr A_{m} -
L_{m}\cdot\mathscr A_{n} & = (n-m)\mathscr A_{n+m}\\ 
\label{WZcc3} L_{n}\cdot \mathscr B_{m} -
J_{m}\cdot\mathscr A_{n} & = (n-m)\mathscr B_{n+m}\\
\label{WZcc4}J_{n}\cdot \mathscr B_{m} -
J_{m}\cdot\mathscr B_{n} & = 0
\end{align}
associated with the algebra \eqref{pertalg}.

It is convenient to use operator-valued generating functions for the
$L_{n}$ and $J_{n}$,
\be\label{LJzdef} L(z) = \sum_{n\geq -1}\frac{L_{n}}{z^{n+2}}\,\cvp
\quad
J(z) = \sum_{n\geq -1}\frac{J_{n}}{z^{n+2}}\,\cdotp\ee
These operators generate anomaly polynomials that can be written
elegantly in terms of the generating functions $R$ and $S$ for the
$u_{n}$s and $v_{n}$s,
\begin{align}\label{Ano1}\mathscr A(z)=\sum_{n\geq -1}\frac{\mathscr
A_{n}}{z^{n+2}}
& = -NW'(z)R(z) + 2 R(z)S(z) + N^{2}\Delta_{R}(z)\, ,\\ 
\label{Ano2}\mathscr B(z)=\sum_{n\geq -1}\frac{\mathscr
B_{n}}{z^{n+2}}
&= -NW'(z)S(z) + S(z)^{2}+N^{2}\Delta_{S}(z)\, ,
\end{align}
where $\Delta_{R}$ and $\Delta_{S}$ are polynomials chosen to cancel
the terms of positive powers in $z$ in the right-hand sides of
\eqref{Ano1} and \eqref{Ano2}.

\subsection{Non-perturbative subtleties and finite $N$}
\subsubsection{The non-perturbative anomaly conjecture}\label{anoconj}

The anomaly polynomials \eqref{ano1} and \eqref{ano2} must vanish
on-shell. The resulting equations are very similar to the planar loop
equations of the one-matrix model, and this hints at the formulation
in terms of the matrix model in \cite{DV}. However, there is a very
important difference with the matrix model, that has been overlooked
in most of the literature, but which was emphasized in
\cite{ferchiral}. In the gauge theory, the number of colors $N$ is
\emph{finite}, and thus the variables that enter in \eqref{ano1} and
\eqref{ano2} are not independent. Actually, only $u_{1},\ldots,u_{N}$
and $v_{0},\ldots,v_{N-1}$ can be independent, all the other
observables being expressed as polynomials in these basic variables.
For example, because $X$ is a $N\times N$ matrix, we have
\be\label{relucl} u_{N+p} = \mathscr
P_{\text{cl},\,p}(u_{1},\ldots,u_{N})\, ,\ p\geq 1\, ,\ee
for some homogeneous polynomials $\mathscr P_{p}$ of degree $N+p$
($u_{n}$ being of degree $n$) that can be easily computed. It is
straightforward to check that the vanishing of the anomaly polynomials
can be consistent with \eqref{relucl} only if the expectation values
do not get quantum corrections at all, providing a proof of the
standard perturbative non-renormalization theorem.

These remarks clearly show that the anomaly polynomials must get
non-per\-tur\-ba\-ti\-ve corrections to be consistent with the
non-trivial non-perturbative corrections to the chiral operators
expectation values \cite{ferchiral}. The precise conjecture about the
anomaly equations can then be stated as follows
\cite{ferchiral}:\smallskip\\
\textsc{Non-perturbative anomaly conjecture}: \emph{The
non-perturbative corrections to \eqref{ano1} and \eqref{ano2} are such
that they can be absorbed in a non-perturbative redefinition of the
variables that enter the equations.}\smallskip\\ 
\noindent This means that, at the expense of \emph{defining} the
variables $u_{n}$ and $v_{n-1}$ for $n>N$ in a suitable way, we can
assume that the anomaly polynomials \eqref{ano1} and \eqref{ano2} are
exact at the non-perturbative level. The only constraints on the
possible definitions of the variables come from the classical limit
and the symmetries of the theory, the $\u_{\text R}$ symmetry
\eqref{charges} as well as the $\u_{\text A}$ symmetry for which the
relevant charges are given by
\be\label{chargeuRA}
\begin{matrix}
& u_{n} & v_{n} & g_{k} & q \\
\uR & 0 & 2 & 2 & 0 \\
\u_{\text A} & n & n & -k-1 & \hphantom{,\,} 2N \, .
\end{matrix}\ee
For example, the $u_{N+p}$ that enter in the anomaly polynomials could
be given by any formula of the form
\be\label{relu} u_{N+p} = \mathscr
P_{p}(u_{1},\ldots,u_{N};q)\, ,\ p\geq 1\, ,\ee
for polynomials $\mathscr P_{p}$ of $\u_{\text A}$ charge $N+p$ that
goes to $\mathscr P_{\text{cl},\, p}$ when $q$ goes to zero. The
precise form of the polynomials $\mathscr P_{p}$ are unknown a priori.
However, a little thinking shows that it is actually quite miraculous
that the vanishing of the anomaly polynomials can be consistent at all
with the existence of non-trivial quantum corrections and relations
like \eqref{relu}. It was then conjectured in \cite{ferchiral} that
the form of the polynomials were actually fixed uniquely by
consistency with the anomaly equations, and that this requirement was
actually equivalent to the extremization of the Dijkgraaf-Vafa
superpotential. This conjecture can be proven, including when flavors
are added to the theory \cite{pchiral}.

In a given non-perturbative microscopic setting, where all the
operators $u_{n}$ and $v_{n}$ are well-defined, the relations like
\eqref{relu} must be fixed. Let us emphasize again that these
relations are mere definitions of what we mean by $u_{N+p}$ for $p\geq
1$, and thus have no dynamical content. In particular, they must be
valid off-shell. In our framework, based on the non-commutative
regularization of the instanton moduli space, we thus expect to find
some explicit form for the polynomials $\mathscr P_{p}$, with
relations \eqref{relu} valid for any values of the boundary conditions
$\a$. This can be easily checked as follows \cite{ferchiral}.

Let us introduce the correlator
\be\label{Fdef} F(z;\a,q) = \vevab{\det (z-X)}\, .\ee
We have
\be\label{RFrel}\frac{F'(z)}{F(z)} = R(z)\, ,\ee
and Nekrasov's formula \eqref{Rform} then implies that
\be\label{Fform} F(z;\a,q) = \frac{1}{2}\Bigl( P(z) + \sqrt{P(z)^{2} -
4 q}\Bigr)\, .\ee
The function $F$ is thus a well-defined meromorphic function on the
curve \eqref{SWcurve}, and in particular it satisfies an algebraic
equation that can be conveniently written in the form
\be\label{Feq} F(z) + \frac{q}{F(z)} = P(z)\, .\ee
Expanding at large $z$, using the fact that
\be\label{Fexp} F(z) = z^{N}e^{-\sum_{n\geq 1}u_{n}/(nz^{n})}\ee
and that all the terms with negative powers of $z$ in the left hand
side of \eqref{Feq} must vanish, we obtain an infinite set of
equations that generate recursively \emph{and are equivalent to} a
specific form for the relations \eqref{relu}. For example, we find
that
\be\label{exCRrel} \mathscr P_{p} = \mathscr P_{\text{cl}\, ,p}\quad
\text{for}\ 1\leq p\leq N-1\, ,\quad \mathscr P_{N} =\mathscr
P_{\text{cl},\, N} + 2Nq\, ,\quad \text{etc}\ldots\ee
This is equivalent to saying that the equation \eqref{Feq} is not
dynamical but simply encodes the off-shell kinematical relations
\eqref{relu} (only the explicit form of the polynomial $P$ is
dynamical). It is extremely tempting to believe that this natural
definition of the operators is precisely the one for which the anomaly
equations take the simple forms \eqref{ano1} and \eqref{ano2}. This is
suggested by all the known results on the theory, and we will check it
explicitly up to two instantons below and to all orders in
\cite{mic3}. However, having non-trivial $q$-dependent relations like
\eqref{exCRrel} between the operators imply some very drastic
consequences on the generators $L_{n}$ and $J_{n}$ that were defined
in perturbation theory by \eqref{varop} or equivalently by
\eqref{LJact}, as we are now going to discuss.

\subsubsection{On the quantum corrected operators $L_{n}$ and $J_{n}$}
\label{qco}

At the non-perturbative level, the operators $L_{n}$ and $J_{n}$
clearly can get quantum corrections for $n\geq 1$ because the
associated transformations are non-linear. This is a well-known field
theoretic effect, that plays a r\^ole in many instances, for example
in the BRST renormalization theory of Yang-Mills: non-linear
transformation rules can be renormalized. Here we are dealing with a
particularly interesting non-perturbative example of this effect.

An obvious question to ask is what kind of quantum corrections can
modify the operators $L_{n}$ and $J_{n}$ and their algebra. This is
important for example if one wish to study the possible
non-perturbative corrections to the anomaly equations by using the
Wess-Zumino consistency conditions, as suggested in \cite{CDSW}. A
natural, albeit na\"\i ve, guess is that the corrections are mild
enough for the operators to remain derivations acting in a closed form
on the chiral ring. For example, focusing on the operators $L_{n}$ and
variables $u_{m}$, we might assume that in the full quantum theory the
most general possibility is to have relations like
\be\label{NPact} L_{n}\cdot u_{m} = -m u_{n+m} + \sum_{k\geq 1}q^{k}
r^{(k)}_{n,m}\ee
and
\be\label{algQ} [L_{n},L_{m}] = (n-m)L_{n+m} + \sum_{k\geq
1}q^{k}L^{(k)}_{n,m}\, ,\ee
where the $r^{(k)}_{n,m}$ are polynomials in the $u_{p}$s and the
$L^{(k)}_{n,m}$ are operators of A-charges $n+m-2Nk$, consistently
with \eqref{chargeuRA}. Note that the constraints on the A-charges
imply that the instanton series in \eqref{NPact} and \eqref{algQ} have
only a finite number of terms. Constraints like \eqref{NPact} are at
the basis of the analysis in \cite{svrcek} for
example.\footnote{Several assumptions and derivations in \cite{svrcek}
are inconsistent and we do not agree with most of the statements in
this paper.} However, and perhaps surprisingly, it turns out that the
non-perturbative quantum corrections to the operators $L_{n}$ and
$J_{n}$ must be much stronger. Actually, the formulas \eqref{NPact}
and \eqref{algQ} are \emph{inconsistent} with the existence of the
quantum corrected relations \eqref{relu}!

The precise statement is as follows:\smallskip\\
\emph{Assume that the anomaly equations are given by \eqref{ano1} and
\eqref{ano2} with the $u_{N+p}$ variables defined by \eqref{relu},
where the polynomials $\mathscr P_{p}$ are deduced from
\eqref{Feq}.\footnote{These are the standard claims about the theory,
and we shall be able to provide a full microscopic derivation below
and in \cite{mic3}.} Assume that relations like \eqref{NPact} and
\eqref{algQ} are also valid. Then necessarily $q=0$, i.e.\ the theory
is classical.}\smallskip\\
Let us derive this result in the simple case $N=2$. We have also done
the analysis in the general case, but it is quite tedious and not
necessary for our purposes. It will be enough to consider a tree-level
superpotential of the form $W(z) = \frac{1}{2}mz^{2}$. From
\eqref{NPact} and \eqref{algQ}, we only need the facts that the
$L_{n}\cdot u_{m}$ and $[L_{n},L_{m}]$ (and thus the associated
Wess-Zumino consistency conditions) are not corrected if $n+m<4$, as
well as
\begin{align}\label{act1} L_{0}\cdot u_{4} &= -4 u_{4} + c_{1} q\, ,\\
\label{act2} L_{2}\cdot u_{2} & = -2 u_{4} + c_{2} q\, ,\\
\label{act3} L_{1}\cdot u_{3} & = -3 u_{4} + c_{3} q\, ,
\end{align}
for some numerical constants $c_{1}$, $c_{2}$ and $c_{3}$. These
constants are not independent. From $[L_{2},L_{0}]=2L_{2}$, we deduce
\be\label{vir1} L_{2}\cdot u_{2}=\frac{1}{2}[L_{2},L_{0}]\cdot
u_{2}=\frac{1}{2}L_{2}\cdot (-2u_{2}) - \frac{1}{2}L_{0}\cdot(-2u_{4})
= -L_{2}\cdot u_{2} + L_{0}\cdot u_{4}\, ,\ee
which implies that
\be\label{c2} c_{2} = \frac{c_{1}}{2}\,\cdotp\ee
Similarly, $[L_{2},L_{1}]=L_{3}$ acting on $u_{1}$ yields
\be\label{c3a} L_{1}\cdot u_{3} = L_{3}\cdot u_{1} + L_{2}\cdot
u_{2}\, ,\ee
and $[L_{3},L_{0}]=3L_{3}$ acting on $u_{1}$ yields, by using
\eqref{c3a},
\be\label{c3b} L_{1}\cdot u_{3} = L_{2}\cdot u_{2} + \frac{1}{4}
L_{0}\cdot u_{4}\, .\ee
From \eqref{act1}, \eqref{act2} and \eqref{c2} we thus get
\be\label{c3} c_{3} = \frac{3c_{1}}{4}\,\cdotp\ee
Let us now use the Wess-Zumino consistency conditions \eqref{WZcc1}
for
$(n,m)=(2,0)$. Using the explicit formulas
\be\label{A0A2} \mathscr A_{0} = -2mu_{2} + 4v_{0}\, ,\quad \mathscr
A_{2} = -2mu_{4} + 4 v_{2} + 2 u_{1}v_{1} + 2 u_{2}v_{0}\ee
and \eqref{act1} and \eqref{act2}, a direct calculation shows that
\be\label{WZa} L_{2}\cdot\mathscr A_{0}-L_{0}\cdot\mathscr A_{2} - 2
\mathscr A_{2} = 2m(c_{1}-c_{2})q = 0\, .\ee
Using \eqref{c2} and \eqref{c3}, we deduce that
\be\label{c1c2c3} c_{1}=c_{2}=c_{3}=0\, .\ee
Let us now use \eqref{exCRrel} in the cases $N=2$, $p=1$ and $p=2$,
\begin{align}\label{u3rel}u_{3} &= \mathscr P_{\text{cl},\,
1}(u_{1},u_{2}) = \frac{3}{2}u_{1}u_{2}-\frac{1}{2}u_{1}^{3}\, ,\\
\label{u4rel}u_{4} &= \mathscr P_{\text{cl},\, 2}(u_{1},u_{2}) + 4 q =
u_{1}u_{3}+\frac{1}{2}u_{2}^{2}-\frac{1}{2}u_{1}^{2}u_{2} + 4q \,
.\end{align}
Acting on \eqref{u3rel} with the operator $L_{1}$, and using
\eqref{c1c2c3}, yields
\be\label{L1onu3} L_{1}\cdot u_{3} = -3 u_{4} = L_{1}\cdot\bigl(
\frac{3}{2}u_{1}u_{2}-\frac{1}{2}u_{1}^{3}\bigr) =
-\frac{3}{2}u_{2}^{2} - 3 u_{1}u_{3} + \frac{3}{2} u_{1}^{2}u_{2}\,
.\ee
This is consistent with \eqref{u4rel} only for $q=0$, as was to be
shown.

\subsection{Non-perturbative generators and algebra}\label{npgenalg}

We have seen in the previous subsection that the quantum corrections
to the generators of the anomaly equations must be very strong, and in
particular must violate ansatz like \eqref{NPact} and \eqref{algQ}. It
is then very difficult to guess the general form of the allowed
corrections a priori. In particular, it seems extremely difficult to
try to derive the non-perturbative anomaly conjecture by using the
Wess-Zumino consistency conditions.

On the other hand, in the microscopic framework of the present paper,
it should be possible in principle to provide a full derivation of the
anomaly equations and associated generators and algebra. In our
framework, we are thus seeking differential operators $L_{n}$ and
$J_{n}$, or more conveniently the generating functions $L(z)$ and
$J(z)$ defined in \eqref{LJzdef}, that act on the microscopic
off-shell variables $a_{i}$,
\be\label{LJmic} L(z) =
\sum_{i=1}^{N}\delta_{z}^{L}a_{i}\,\frac{\partial}{\partial
a_{i}}\,\cvp\quad J(z) =
\sum_{i=1}^{N}\delta_{z}^{J}a_{i}\,\frac{\partial}{\partial a_{i}}\,
\cvp\ee
and such that
\begin{align}\label{anomic1}\begin{split}
N L(z)\cdot\wmic(\a,\g,q) & =\mathscr
A(z;\a,\g,q)\\ &=
-NW'(z)R(z;\a,q) + 2 R(z; \a,q)S(z;\a,\g,q)\, ,
\end{split}\\ \label{anomic2}\begin{split}
N J(z)\cdot\wmic(\a,\g,q)  &=\mathscr B(z;\a,\g,q)\\ &=
-NW'(z)S(z;\a,\g,q) + S(z;\a,\g,q)^{2}\, .\end{split}\end{align}
The functions $R(z;\a,q)$ and $S(z;\a,\g,q)$ have been studied
extensively in Sections 2 and 3. $R$ is explicitly known from the
results of \cite{nekrasovb}, see equation \eqref{Rform}. On the other
hand, $S$ can in principle be obtained by summing over colored
partitions from \eqref{vnform}, but we only know its explicit form up
to two instantons from the calculations of Section 3.

There is a very natural proposal for the operators $L(z)$ and $J(z)$. 
We conjecture that
\begin{align}\label{conjL} \delta_{z}^{L}a_{i}& =
\frac{1}{2i\pi}\oint_{\alpha_{i}}\frac{R(z';\a,q)}{z'-z}\,\d z'\, ,\\
\label{conjJ}\delta_{z}^{J}a_{i} &=
\frac{1}{2i\pi}\oint_{\alpha_{i}}\frac{S(z';\a,\g,q)}{z'-z}\,\d z'\,
.\end{align}
In these formulas, the point $z$ is chosen to be outside the contours
$\alpha_{i}$ that were defined in Section \ref{scasec}. For the
$L_{n}$ and $J_{n}$, the corresponding explicit formulas read
\begin{align}\label{conjLn} L_{n}& =
-\frac{1}{2i\pi}\sum_{i=1}^{N}\oint_{\alpha_{i}}z^{n+1}R(z;\a,q)\,\d
z\,\frac{\partial}{\partial a_{i}}\,\cvp\\\label{conjJn} J_{n}& =
-\frac{1}{2i\pi}\sum_{i=1}^{N}\oint_{\alpha_{i}}z^{n+1}S(z;\a,\g,q)\,\d
z\,\frac{\partial}{\partial a_{i}}\,\cdotp\end{align}
We would like to make two comments on the above formulas.

First, it is not obvious a priori that the formulas for $J(z)$ or
$J_{n}$ make sense, because we do not know if $S(z)$ is a well-defined
function on the curve \eqref{SWcurve}. Actually, since the contours
$\alpha_{i}$ lie entirely on the first sheet of the surface, which is
defined by the asymptotic conditions
\be\label{asyRS}
R(z;\a,q)\underset{z\rightarrow\infty}{\sim}\frac{N}{z}\,\cvp\quad
S(z;\a,\g,q)\underset{z\rightarrow\infty}{\sim}\frac{v_{0}(\a,\g,q)}{z}
\,\cvp\ee
all we need is that $S(z)$ is well defined on this first sheet, with
the same branch cuts as $R(z)$. In particular, the conditions
\be\label{condonS}\oint_{\alpha_{i}}\! S'(z;\a,\g,q)\,\d z = 0\ee
must be satisfied. Anticipating a bit the results derived in
\cite{mic3}, it can be shown that $S'(z)$ is a well-defined
meromorphic function on \eqref{SWcurve} satisfying \eqref{condonS},
ensuring that the formulas \eqref{conjJ} and \eqref{conjJn} do make
sense. However, it turns out that the function $S(z)$ itself is
\emph{not} well defined on \eqref{SWcurve}.

The second comment we would like to make is related to the discussion
in Section \ref{qco}. It is actually quite obvious that a formula like
\eqref{conjLn} must violate \eqref{NPact} (with similar statements for
the $J_{n}$). The reason is that $L_{n}\cdot u_{m}(\a,q)$ will in
general be a well-defined function of the $a_{i}$, but a
\emph{multi}-valued function of the $u_{p}$. This is the consequence
of the well-known non-trivial monodromies that the variables $a_{i}$
undergo in the $u_{p}$-space. Similarly, the algebra of the operators
$L_{n}$ and $J_{n}$ defined by \eqref{conjLn} and \eqref{conjJn} is
not closed. This can be checked straightforwardly from \eqref{Rform}
and the formulas in Section 4.1 of \cite{mic1}. In order to obtain a
closed algebra, we need to enlarge the set of operators considerably.
Let us see how this work in the case of the operators $L_{n}$. We set,
for any meromorphic one-form $\omega$ on \eqref{SWcurve},
\be\label{sk} \sigma_{i}(\omega) =
\frac{1}{2i\pi}\oint_{\alpha_{i}}\!\omega\, ,\ee
and associate to $\omega$ the differential operator defined by
\be\label{Lgendef} L(\omega) =
\sum_{i=1}^{N}\sigma_{i}(\omega)\frac{\partial}{\partial
a_{i}}\,\cdotp\ee
The operators $L_{n}$ are of this form,
\be\label{Lningen} L_{n} = L(\omega_{n})\, ,\quad \omega_{n} =
-z^{n+1}R(z)\, \d z\, .\ee
The commutator of two operators $L(\omega)$ and $L(\eta)$ is given
in terms of the skew product
\be\label{skewL} \bigl\langle\omega,\eta\bigr\rangle =
\sum_{i=1}^{N}\Bigl(\sigma_{i}(\omega)\frac{\partial\eta}{\partial
a_{i}} - \sigma_{i}(\eta)\frac{\partial\omega}{\partial
a_{i}}\Bigr)\ee
by
\be\label{Lie} \bigl[ L(\omega),L(\eta)\bigr] =
L(\langle\omega,\eta\rangle)\, .\ee
Taking the derivative of forms with respect to $a_{i}$ can introduce
poles at the branching points $x_{i}^{\pm}$ of the curve
\eqref{SWcurve}. For this reason, the commutators of the $L_{n}$, and
then the commutators of commutators, etc, will generate operators
$L(\omega)$ with forms $\omega$ having poles of higher and higher
orders at the branching points $x_{i}^{\pm}$. The resulting infinite
dimensional algebra is quite interesting and would deserve further
study. In the limit $q\rightarrow 0$ it has the partial Virasoro
algebra as a closed subalgebra.

\subsection{Checks in the instanton expansion}
\subsubsection{The anomaly equations}

Let us now check explicitly \eqref{anomic1} and \eqref{anomic2} by
using the results of Section 3. The calculation is straightforward,
but quite tedious. Actually, finding the correct anomaly polynomials
look like a little miracle in the present formalism. This is very
unlike the case of the matrix model approach, where the anomaly
equations are the most natural identities, and follow directly from
the properties of the matrix integral. In the present microscopic
formalism based on the sum over colored partitions, we do not have
such a simple interpretation.

We have performed all our calculations at the \emph{two}-instantons
order. However, the intermediate formulas are so complicated that we
are simply going to indicate the main steps, writing explicitly only
the terms relevant to the one-instanton order.

First, we write the generating functions explicitly using the formulas
derived in Section 3,
\begin{align}\label{R1inst}& R(z;\a,q)=\sum_i \frac{1}{z-a_i} + 2q
\sum_i \frac{1}{\prod_{l\neq i}a_{il}^2}\frac{1}{(z-a_i)^3}+\mathcal
O(q^2)\, ,\\ 
\label{S1inst}& S(z;\a,\g,q)=N q \sum_i \frac{W''(a_i)}{\prod_{l\neq
i}a_{il}^2}\frac{1}{z-a_i}+\mathcal O(q^2)\, . \end{align}
We see that in the small $q$ expansion, the functions $R$ and $S$ are
meromorphic functions on the complex plane with poles at the points
$z=a_{i}$. This feature is maintained at any finite order in $q$, with
poles of higher and higher orders as the instanton number increases.
The $\alpha_{i}$-periods of differential forms involving $R$ and $S$
thus reduce to a sum over the residues at $a_{i}$. Using
\eqref{conjLn} and \eqref{conjJn}, we can get in this way the explicit
formulas for the operators $L_{n}$ and $J_{n}$,
\begin{align}\label{Ln1}&L_n=-\sum_i \bigg(a_i^{n+1}+q(n+1)n\,
\frac{a_i^{n-1}}{\prod_{l\neq i}a_{il}^2}\bigg)
\frac{\partial}{\partial a_i} +\mathcal O(q^2)\, , \\
\label{Jn1}&J_n=-N q \sum_i \frac{W''(a_i)a_i^{n+1}}{\prod_{l\neq
i}a_{il}^2} \frac{\partial}{\partial a_i} +\mathcal O(q^2)\, .
\end{align}
We need next to compute $\partial\wmic/\partial a_{i}$. From
\eqref{Wmicresa} we know that
\be\label{Wmic1inst}\wmic(\a,\g,q)=\sum_{m\geq
0}\frac{g_m}{m+1}\Big( u_{m+1}^{\cl}(\a) + u_{m+1}^{(1,0)}(\a)\,
q+\mathcal O(q^2) \Big)\, , \ee
from which we find, using \eqref{u0m} and \eqref{oneins0},
\be\label{eom1}\frac{\partial \wmic}{\partial a_i}= W'(a_i) + q
\biggl[ \frac{1}{\prod_{l\neq i}a_{il}^2}\Bigl(W'''(a_i)-2\sum_{l\neq
i}\frac{1}{a_{il}} W''(a_i) \Bigr) -2\sum_{j\neq i}
\frac{W''(a_j)}{\prod_{l\neq
j}a_{jl}^2}\frac{1}{a_{ij}}\biggr]+\mathcal O(q^2)\, . \ee 
Combining \eqref{eom1} with \eqref{Ln1} and \eqref{Jn1}, we can then
check explicitly that
\be\label{anocheck}NL_{n}\cdot\wmic = \mathscr A_{n}+\mathcal
O(q^{2})\, ,\quad NJ_{n}\cdot\wmic = \mathscr B_{n}+\mathcal
O(q^{2})\, .\ee
Repeating the same calculation, but now including all the relevant
two-instantons terms, we have actually explicitly checked, at the cost
of considerable algebra, that
\be\label{anocheck2}NL_{n}\cdot\wmic = \mathscr A_{n}+\mathcal
O(q^{3})\, ,\quad NJ_{n}\cdot\wmic = \mathscr B_{n}+\mathcal
O(q^{3})\, ,\ee
or equivalently that \eqref{anomic1} and \eqref{anomic2} are valid up
to terms of order $q^{3}$.

Note that the above results immediately imply that the microscopic
approach match the Dijkgraaf-Vafa approach, at least up to two
instantons. Indeed, when the equations \eqref{qem} are satisfied, we
automatically get
\be\label{qemano} NL(z)\cdot\wmic = 0 = \mathscr A(z)\, ,\quad 
NJ(z)\cdot\wmic = 0 =\mathscr B(z)\, .\ee
In the Dijkgraaf-Vafa formalism, these equations must be supplemented
by the extremization of the glueball superpotential. However, it is
well-known (see for example \cite{ferproof,ferchiral}) that this is
equivalent to the fact that the quantum characteristic function
\eqref{Fdef} satisfies the algebraic equation \eqref{Feq}. This latter
equation is automatically implemented in the microscopic approach.

There is, of course, a limitation in working at a finite order in the
instanton expansion. The equations of motion \eqref{qem} then allow to
study only the Coulomb vacuum of the theory, in which the unbroken
gauge group has only $\u$ factors. This limitation will be waived in
\cite{mic3}, using the results of \cite{mic1}, by providing an exact
analysis independent of the small $q$ approximation.

\subsubsection{The algebra}

Let us now compute the first non-trivial quantum corrections to the
perturbative algebra \eqref{pertalg}. From \eqref{Ln1} we find
\be\label{LnLm2}\begin{split} [L_n,&L_m] = (n-m)
L_{n+m}\\ & +\frac{2q}{\prod_{l\neq i}a_{il}^2}
\Bigl(n(n+1)\, a_i^{n-1}\sum_j \sum_{q_1+q_2=m} a_i^{q_1}
a_j^{q_2} - \bigl(n\leftrightarrow m\bigr)
\Bigr)\frac{\partial}{\partial a_{i}} +\mathcal O(q^2)\, . 
\end{split}\ee
Similarly, using \eqref{Jn1} we find
\be\label{LnJm}\begin{split}[L_n,&J_m] = (n-m) J_{n+m}\\& + \frac{N
q}{\prod_{l\neq i}a_{il}^2} \bigg(W'''(a_i)\, a_i^{n+m+2} -2W''(a_i)\,
a_i^{m+1} \sum_{j\neq i} \frac{a_i^{n+1}-a_j^{n+1}}{a_{ij}} \bigg)
\frac{\partial}{\partial a_{i}} 
+\mathcal O(q^2)\end{split}\ee
and
\be\label{JnJm}\begin{split}[J_n,J_m] = &N^2 q^2 \biggl[
\bigl(m-n\bigr) \frac{W''(a_i)^2\, a_i^{n+m+1}}{\prod_{l\neq
i}a_{il}^4} \\ &\hskip 1cm + \frac{2 W''(a_i)}{\prod_{l\neq
i}a_{il}^2} \Bigr(a_i^{m+1} \sum_{j\neq
i}\frac{W''(a_j)a_j^{n+1}}{a_{ij}\prod_{l\neq j}a_{jl}^2} - \bigl( n
\leftrightarrow m \bigr) \Bigr)\biggr] +\mathcal O(q^3)\,
.\end{split}\ee
An interesting feature of the above equations is to show explicitly
that the algebra does not close, as discussed in \ref{npgenalg}: the
quantum corrections would have to be linear combinations of the
operators at lower order, which is impossible due to the pole
structure.

\section{Outlook}
\setcounter{equation}{0}

In this paper, following \cite{mic1}, we have provided a detailed
microscopic analysis of the $\nn=1$ gauge theory with one adjoint
chiral multiplet and arbitrary tree-level superpotential. We have
shown how to use Nekrasov's instanton technology to derive many deep
results in $\nn=1$ gauge theories. In particular, we have provided the
first non-perturbative discussion of the generalized Konishi anomaly
equations, putting forward the subtle constraints coming from working
at finite $N$ and deriving the strong quantum corrections to the
operators that generate them. We have also computed explicitly the
first two terms in the instanton expansion of various operators in the
$\Omega$-background, including the generating function $S(z;\a,\g,q)$
for the generalized glueball operators.

Our calculations were limited to the two-instantons order. A full
solution of the problem, which includes in particular the calculation
of the function $S$ and the derivation of the equations
\eqref{anomic1} and \eqref{anomic2} is of course highly desirable. It
will be presented in a forthcoming publication \cite{mic3}. The fact
that the present microscopic formalism, based on the sum over colored
partitions, can match the results from the matrix model approach is a
very deep property, clearly related to the open/closed string duality.

It would also be extremely interesting to study the theory with
flavors of fundamental quarks and other models with various gauge
groups and matter contents along the same line. It seems that the
derivation, from a direct microscopic analysis, of all the conjectured
exact results in $\nn=1$ gauge theories is now at hand. After almost
fifteen years of intense study of the non-perturbative properties of
these theories, we believe that this is a highly satisfactory result.

\subsection*{Acknowledgements}

This work is supported in part by the belgian Fonds de la Recherche
Fondamentale Collective (grant 2.4655.07), the belgian Institut
Interuniversitaire des Sciences Nucl\'eaires (grant 4.4505.86), the
Interuniversity Attraction Poles Programme (Belgian Science Policy)
and by the European Commission FP6 programme MRTN-CT-2004-005104 (in
association with V.\ U.\ Brussels). Vincent Wens is a junior
researcher (Aspirant) at the belgian Fonds National de la Recherche
Scientifique. Frank Ferrari is on leave of absence from the Centre
National de la Recherche Scientifique, Laboratoire de Physique
Th\'eorique de l'\'Ecole Normale Sup\'erieure, Paris, France.

\renewcommand{\thesection}{\Alph{section}}
\renewcommand{\thesubsection}{\arabic{subsection}}
\renewcommand{\theequation}{A.\arabic{equation}}
\setcounter{section}{0}
\section*{Appendix}

In this appendix, we prove the equivalence between the formulas
\eqref{mukgen} and \eqref{mukgenbis} for the measure on the set of
colored partitions. Both formulas have appeared in the literature,
starting from \cite{nekrasova}, but often in erroneous or
undeterminate forms (for example by writing them in terms of ambiguous
infinite products). Since having the exact formulas was essential to 
perform our explicit calculations, we have been extremely careful in
deriving them and we hope that this appendix will clarify the main
properties of the measure factor.

We shall need the following simple\smallskip\\
\textsc{Lemma}: \emph{Let $\mathsf k$ be a partition and $z \in
\mathbb C$. Then}
\be\label{lemma}\frac{1}{z-k_1}\prod_{\beta=1}^{k_1}\frac{z+\tilde
k_{\beta}-\beta}{z+\tilde k_{\beta}-\beta+1} = \frac{1}{z+\tilde
k_1}\prod_{\alpha=1}^{\tilde k_1}
\frac{z+\alpha-k_{\alpha}}{z+\alpha-k_{\alpha}-1}\, \cdotp \ee
The proof is made recursively on the number of columns of the
partition $\mathsf k$. We first consider a partition whose Young
tableau $Y_{\mathsf k}$ has a single column of arbitrary length, i.e.\
$k_{\alpha} = 1$ for $1 \leq \alpha \leq \tilde k_1$. In this case,
the left hand side of \eqref{lemma} reads
\be\label{startind1} \frac{1}{z-1}\frac{z+\tilde k_1-1}{z+\tilde
k_1}\, \cvp \ee
consistently with the right hand side which, using the many
cancellations between the numerator and the denominator in the
product, reads
\be\label{startind2} \frac{1}{z+\tilde k_1} \prod_{\alpha=1}^{\tilde
k_1} \frac{z+\alpha-1}{z+\alpha-2} = \frac{1}{z+\tilde
k_1}\frac{z+\tilde k_1-1}{z-1}\, \cdotp\ee
Now, we assume that the lemma is true for partitions $\mathsf k$ with
$k_{1}$ columns in the Young tableau. Let us consider a partition
$\mathsf k'$ with $k_{1}'=k_{1}+1$ columns. Its Young tableau
$Y_{\mathsf k'}$ can be built by adding its first column to a Young
tableau $Y_{\mathsf k}$ having only $k_{1}$ columns. Precisely, we
have $k_{\alpha}' = k_{\alpha}+1$ for $1 \leq \alpha \leq \tilde k_1$
and $k_{\alpha}'=1$ for $\tilde k_1+1 \leq \alpha \leq \tilde k_1'$.
The left hand side of \eqref{lemma} for $\mathsf k'$ is
\be\label{ind1}\begin{split}
\frac{1}{z-k_1'}\prod_{\beta=1}^{k_1'}\frac{z+\tilde
k_{\beta}'-\beta}{z+\tilde k_{\beta}'-\beta+1} &=
\frac{1}{z-k_1-1}\prod_{\beta=1}^{k_1+1}\frac{z+\tilde
k_{\beta}'-\beta}{z+\tilde k_{\beta}'-\beta+1} \\ &=
\frac{1}{z-k_1-1}\frac{z+\tilde k_1'-1}{z+\tilde k_1'}
\prod_{\beta=1}^{k_1}\frac{z+\tilde k_{\beta}-\beta-1}{z+\tilde
k_{\beta}-\beta} \, \cdotp \end{split}\ee
In the second line of \eqref{ind1} we have explicitly splitted the
product over $\beta$ into the term $\beta=1$ and the product over
$2\leq \beta \leq k_1+1$ for which we can use $\tilde
k_{\beta}'=\tilde k_{\beta-1}$. Using the recursion hypothesis for
$\mathsf k$ with $z-1$ replacing $z$, we can compute the product over 
$\beta$ in the second line of \eqref{ind1}, which yields
\be\label{ind2} \frac{1}{z-k_1'}\prod_{\beta=1}^{k_1'}\frac{z+\tilde
k_{\beta}'-\beta}{z+\tilde k_{\beta}'-\beta+1} = \frac{z+\tilde
k_1'-1}{z+\tilde k_1'}\frac{1}{z+\tilde k_1 - 1}
\prod_{\alpha=1}^{\tilde k_1}
\frac{z+\alpha-k_{\alpha}-1}{z+\alpha-k_{\alpha}-2} \, \cdotp \ee
On the other hand, we compute the right hand side of \eqref{lemma} for
$\mathsf k'$ by splitting the product over $\alpha$ into two terms as
\be\label{ind3} \frac{1}{z+\tilde
k_1'}\prod_{\alpha=1}^{\tilde k_1'}
\frac{z+\alpha-k_{\alpha}'}{z+\alpha-k_{\alpha}'-1} =
\frac{1}{z+\tilde k_1'}\prod_{\alpha=1}^{\tilde k_1}
\frac{z+\alpha-k_{\alpha}-1}{z+\alpha-k_{\alpha}-2}
\prod_{\alpha=\tilde k_1+1}^{\tilde k_1'}
\frac{z+\alpha-1}{z+\alpha-2}\,\cdotp\ee
Using the many cancellations in the above products, we find
\be\label{ind4} \frac{1}{z+\tilde k_1'}\prod_{\alpha=1}^{\tilde k_1'}
\frac{z+\alpha-k_{\alpha}'}{z+\alpha-k_{\alpha}'-1} =
\frac{1}{z+\tilde k_1'}\prod_{\alpha=1}^{\tilde k_1}
\frac{z+\alpha-k_{\alpha}-1}{z+\alpha-k_{\alpha}-2} \, \frac{z+\tilde
k_1'-1}{z+\tilde k_1-1} \, \cvp\ee
matching with \eqref{ind2}, which proves the lemma.

A useful corrolary of \eqref{lemma} is that, for any integer $K\geq
0$,
\be\label{lemmabis}
\prod_{\beta=1}^{k_1}\frac{z+\tilde k_{\beta}-\beta}{z+\tilde
k_{\beta}-\beta+K} =
\prod_{\beta'=1}^K\frac{z+\beta'-1-k_1}{z+\beta'-1+\tilde k_1}\,
\prod_{\alpha=1}^{\tilde
k_1}\frac{z+\alpha-k_{\alpha}+K-1}{z+\alpha-k_{\alpha}-1}\, \cdotp
\ee
This identity is very useful to relate products over the columns of a
Young tableau to products over the rows of the same tableau, which is
exactly what is needed to go from \eqref{mukgen} to \eqref{mukgenbis}.
Using the notation \eqref{aijdef}, let us rewrite \eqref{mukgen} and
\eqref{mukgenbis} as
\be\label{mgB} \mu_{\cpart} = \prod_{i=1}^{N}\mu_{\mathsf k_{i}}
\prod_{i<j} \nu_{\cpart,\, ij} =\prod_{i=1}^{N}\mu_{\mathsf k_{i}}
\prod_{i<j} \kappa_{\cpart,\, ij}\ee
with
\begin{align}\label{mbB2}&\begin{aligned} \nu_{\cpart,\, ij}=&
\prod_{\Box_{(\alpha,\beta)}\in Y_{\mathsf
k_{i}}}\frac{1}{a_{ij}+\eps(\beta-\alpha)}\prod_{\Box_{(\alpha,\beta)}\in
Y_{\mathsf k_{j}}}\frac{-1}{a_{ij}+\eps(\alpha-\beta)}\ \times \\&
\prod_{\alpha=1}^{\tilde k_{i,1}}\prod_{\beta=1}^{k_{j,1}}
\frac{\bigl( a_{ij}+\eps (\tilde k_{j,\beta}-\alpha-\beta
+1)\bigr)\bigl(a_{ij}+\eps(k_{i,\alpha}-\beta-\alpha+1)\bigr)}
{\bigl(a_{ij}+\eps(1-\alpha-\beta)\bigr) \bigl(a_{ij}+\eps(\tilde
k_{j,\beta}-\alpha+ k_{i,\alpha}-\beta +1)\bigr)}\,
\cvp\end{aligned}\\ \label{mbB2bis} &\begin{aligned} \kappa_{\cpart,\,
ij}=&(-1)^{|\mathsf k_{j}|} \prod_{\alpha_{1}=1}^{\tilde k_{i,1}}
\prod_{\alpha_{2}=1}^{\tilde k_{j,1}} \frac{
a_{ij}+\eps(k_{i,\alpha_{1}}-k_{j,\alpha_{2}}
-\alpha_{1}+\alpha_{2})}{a_{ij}+
\eps(\alpha_{2}-\alpha_{1})}\times\\& \prod_{\Box_{(\alpha,\beta)}\in
Y_{\mathsf k_{i}}} \frac{1}{a_{ij}+\eps(\beta-\alpha + \tilde
k_{j,1})} \prod_{\Box_{(\alpha,\beta)}\in Y_{\mathsf k_{j}}}
\frac{1}{a_{ij}-\eps(\beta-\alpha + \tilde
k_{i,1})}\,\cdotp\end{aligned}
\end{align}
We claim that 
\be\label{claimnu} \nu_{\cpart,\, ij} = \kappa_{\cpart,\, ij}\, ,\ee
which is a slightly stronger result that the equality between
\eqref{mukgen} and \eqref{mukgenbis}. To prove this claim, we use
\eqref{lemmabis} for the partition $\mathsf k_j$, with
$K=k_{i,\alpha}$ and $z=a_{ij}/\eps -\alpha+1$. This yields
\be\label{reshuffl1}\begin{split}&\prod_{\alpha=1}^{\tilde
k_{i,1}}\prod_{\beta=1}^{k_{j,1}}\frac{a_{ij}+\eps (\tilde
k_{j,\beta}-\alpha-\beta+1)}{a_{ij}+\eps(\tilde k_{j,\beta}-\alpha+
k_{i,\alpha}-\beta+1)} =\\&\hskip 1cm \prod_{\Box_{(\alpha,\beta)}\in
Y_{\mathsf k_{i}}} \frac{a_{ij}+\eps (\beta -\alpha
-k_{j,1})}{a_{ij}+\eps (\beta -\alpha +\tilde k_{j,1})}\, \times
\prod_{\alpha=1}^{\tilde k_{i,1}}\prod_{\alpha'=1}^{\tilde
k_{j,1}}\frac{a_{ij}+\eps
(k_{i,\alpha}-k_{j,\alpha'}-\alpha+\alpha')}{a_{ij}+\eps
(-k_{j,\alpha'}-\alpha+\alpha')}\, \cdotp \end{split}\ee
Moreover, it is straightforward to check the following identities,
that are obtained using the many cancellations between the numerators
and the denominators in the right hand side of the equations,
\begin{align}\label{reshuffl2}\begin{aligned}\prod_{\alpha=1}^{\tilde
k_{i,1}}\prod_{\beta=1}^{k_{j,1}}
\frac{a_{ij}+\eps(k_{i,\alpha}-\beta-\alpha+1)}{a_{ij}+
\eps(1-\beta-\alpha)}= &\prod_{\alpha=1}^{\tilde
k_{i,1}}\prod_{\beta=1}^{k_{j,1}}\prod_{\beta'=1}^{k_{i,\alpha}}
\frac{a_{ij}+\eps(\beta'-\alpha-\beta+1)}{a_{ij}+
\eps(\beta'-\alpha-\beta)}\\
= & \prod_{\Box_{(\alpha,\beta)}\in Y_{\mathsf k_{i}}}
\frac{a_{ij}+\eps(\beta-\alpha)}{a_{ij}+\eps(\beta-\alpha-k_{j,1})}\,
\cvp\end{aligned}\\
\label{reshuffl3}\begin{aligned}\prod_{\alpha=1}^{\tilde
k_{i,1}}\prod_{\alpha'=1}^{\tilde
k_{j,1}}\frac{a_{ij}+\eps(\alpha'-\alpha)}{a_{ij}+
\eps(\alpha'-\alpha-k_{j,\alpha'})}&= \prod_{\alpha=1}^{\tilde
k_{i,1}}\prod_{\alpha'=1}^{\tilde
k_{j,1}}\prod_{\beta=1}^{k_{j,\alpha'}}
\frac{a_{ij}+\eps(\alpha'-\alpha-\beta+1)}{a_{ij}+
\eps(\alpha'-\alpha-\beta)}\\ &= \prod_{\Box_{(\alpha,\beta)}\in
Y_{\mathsf k_{j}}}
\frac{a_{ij}+\eps(\alpha-\beta)}{a_{ij}+\eps(\alpha-\beta-\tilde
k_{i,1})}\, \cdotp\end{aligned} \end{align}
Using \eqref{reshuffl1}, \eqref{reshuffl2} and \eqref{reshuffl3} in
\eqref{mbB2}, we find \eqref{claimnu} as we wished.

Let us note that the square of the formula \eqref{mukgenbis} can be
written elegantly as follows,
\begin{multline}\label{mukgensquare} \mu_{\cpart}^2
= (-1)^{N |\cpart |} \prod_{i,j}\Biggl[\prod_{\alpha_{1}=1}^{\tilde
k_{i,1}}
\prod_{\alpha_{2}=1}^{\tilde k_{j,1}} \frac{
a_{ij}+\eps(k_{i,\alpha_{1}}-k_{j,\alpha_{2}}
-\alpha_{1}+\alpha_{2})}{a_{ij}+
\eps(\alpha_{2}-\alpha_{1})}\\ 
\prod_{\Box_{(\alpha,\beta)}\in Y_{\mathsf k_{i}}}
\frac{1}{a_{ij}+\eps(\beta-\alpha + \tilde k_{j,1})}
\prod_{\Box_{(\alpha,\beta)}\in Y_{\mathsf k_{j}}}
\frac{1}{a_{ij}-\eps(\beta-\alpha + \tilde k_{i,1})}\Biggr]\, \cvp
\end{multline}
with the rule that the ill-defined terms corresponding to $i=j$ and
$\alpha_1=\alpha_2$ in \eqref{mukgensquare} are left out. In this
form, the analogy with the $N=1$ case \eqref{mukN1bis}, as well as the
permutation symmetry \eqref{perm}, are obvious.

Let us use the previous results to show that
\be\label{mukparity} \mu_{\vec{\tilde{\mathsf k}}}(\a,\eps) =
\mu_{\cpart}(\a,-\eps) \, , \ee
where $\vec{\tilde{\mathsf k}}$ is the colored partition dual to
$\cpart$. This is shown in two steps. First, from the explicit
expression \eqref{mbB2}, it is clear that
\be\label{idA1} \nu_{\vec{\tilde{\mathsf k}},\,ij}(\a,\eps) = 
\nu_{\cpart,\, ji}(\a,-\eps)\, .\ee
Using \eqref{claimnu}, this is equivalent to
\be\label{idA2} \kappa_{\vec{\tilde{\mathsf k}},\,ij}(\a,\eps) = 
\kappa_{\cpart,\, ji}(\a,-\eps)\, .\ee
Now, it is immediate to check from \eqref{mbB2bis} that
\be\label{idA3} \kappa_{\cpart,\, ij} = \kappa_{\cpart,\, ji}\, ,\ee
and thus
\be\label{idA4} \kappa_{\vec{\tilde{\mathsf k}},\,ij}(\a,\eps) =
\kappa_{\cpart,\, ij}(\a,-\eps)\, .\ee
Equation \eqref{mukparity} then immediately follows from \eqref{mgB}.

This implies that the partition function \eqref{partfunc} is an even
function of $\eps$, because the sum of the contributions from a given
colored partition and its dual will have this property,
\be\label{Zparity}Z_{\eps}(\a,q,\eps) = Z_{\eps}(\a,q,-\eps)\, . \ee
Moreover, it can also be shown straightforwardly, doing with sums what
we have done with products in \eqref{reshuffl2} and \eqref{reshuffl3},
that equation \eqref{unmic} can be rewritten in the form
\begin{multline}\label{unmicalt} u_{n,\cpart} = \sum_{i=1}^{N}\biggl[
a_{i}^{n} + \sum_{\beta=1}^{k_{i,1}}\Bigl(\bigl( a_{i} -
\eps(\tilde{k}_{i,\beta}-\beta+1)\bigr)^{n} - \bigl(a_{i}-\eps
(\tilde{k}_{i,\beta}-\beta)\bigr)^{n}\\ + \bigl( a_{i} +
\eps\beta\bigr)^{n}
- \bigl( a_{i}+\eps(\beta-1)\bigr)^{n}\Bigr)\biggr]\, .\end{multline}
This implies that
\be\label{unkparity} u_{n,\cpart}(\a,\eps)=u_{n,\vec{\tilde{\mathsf
k}}}(\a,-\eps) \, . \ee
Combining \eqref{mukparity} and \eqref{unkparity}, we see that
correlators built from the scalar operators, which include the
glueballs \eqref{vnform}, are even functions of $\eps$. This is
non-trivial in the colored partition formalism, but this property must
clearly be true in view of the definition \eqref{Omegamat}.

\end{document}